\documentclass[letterpaper,11pt]{article}
\pdfoutput=1
\usepackage{jheppub}
\usepackage{slashed}
\usepackage{graphicx}
\usepackage{subfigure}
\usepackage{soul}
\usepackage{amsmath}
\usepackage{multirow}
\usepackage{comment}
\usepackage{hyperref}
\usepackage{epstopdf}
\usepackage{datetime}

\def\CO{\mathcal{O}}

\def\gev{\,{\rm GeV}}

\def\to{\rightarrow}

\def\ha{\widehat{\alpha}}

\newcommand{\lsim}{\mathrel{\mathop{\kern 0pt \rlap
  {\raise.2ex\hbox{$<$}}}
  \lower.9ex\hbox{\kern-.190em $\sim$}}}
\newcommand{\gsim}{\mathrel{\mathop{\kern 0pt \rlap
  {\raise.2ex\hbox{$>$}}}
  \lower.9ex\hbox{\kern-.190em $\sim$}}}

\definecolor{pink}{RGB}{255,105,180}

\def\abi{\,{\rm ab}^{-1}}

\newcommand{\ee}{{\ensuremath{e^{+} e^{-}}}}
\newcommand{\mm}{{\ensuremath{\mu^{+} \mu^{-}}}}

\begin{document}

\title{Beyond Higgs Couplings: Probing the Higgs with Angular Observables at Future $e^+ e^-$ Colliders}

\author[a]{Nathaniel Craig\footnote{Email: ncraig@physics.ucsb.edu}}
\author[b]{, Jiayin Gu\footnote{Email: gujy@ihep.ac.cn}}
\author[c]{, Zhen Liu\footnote{Email: zliu2@fnal.gov}}
\author[b]{, Kechen Wang\footnote{Email: kechen@ihep.ac.cn}}

\affiliation[a]{Department of Physics, University of California, Santa Barbara, CA 93106, U.S.A.}
\affiliation[b]{Center for Future High Energy Physics, Institute of High Energy Physics, \\ Chinese Academy of Sciences, Beijing 100049, China}
\affiliation[c]{Theoretical Physics Department, Fermi National Accelerator Laboratory, Batavia, IL 60510, U.S.A.}

\abstract{We study angular observables in the $\ee\to Z H\to \ell^+ \ell^-\,b\bar{b}$ channel at future circular $e^+ e^-$ colliders such as CEPC and FCC-ee. Taking into account the impact of realistic cut acceptance and detector effects, we forecast the precision of six angular asymmetries at CEPC (FCC-ee) with center-of-mass energy $\sqrt{s} =$ 240 GeV and 5 (30) ${\rm ab}^{-1}$ integrated luminosity. We then determine the projected sensitivity to a range of operators relevant for the Higgs-strahlung process in the dimension-6 Higgs EFT. Our results show that angular observables provide complementary sensitivity to rate measurements when constraining various tensor structures arising from new physics. We further find that angular asymmetries provide a novel means of both probing BSM corrections to the $H Z \gamma$ coupling and constraining the ``blind spot'' in indirect limits on supersymmetric scalar top partners.  }


\preprint{
\begin{flushright}
FERMILAB-PUB-15-569-T
\end{flushright}
}

\maketitle
\flushbottom

\section{Introduction}

Following the discovery of a Standard Model-like Higgs at the LHC \cite{Aad:2012tfa, Chatrchyan:2012xdj}, the study of Higgs properties has become one of the highest priorities for current and future colliders. High-luminosity electron-positron colliders are particularly well suited to this end, promising a large sample of relatively clean Higgs production events and the ability to directly probe Higgs properties in a model-independent fashion. Such precision tests of Higgs couplings will provide a window into physics beyond the Standard Model (BSM) well above the weak scale.

Thus far much attention has focused on the potential of future $e^+ e^-$ colliders to probe deviations in Higgs properties in terms of a re-scaling of Standard Model couplings \cite{Asner:2013psa,Gomez-Ceballos:2013zzn,CEPCPreCDR}, with sensitivity exceeding the percent level in some channels. However, in general deviations in Higgs properties may encode additional information, for example in the form of operators with different tensor structure in the Higgs Effective Field Theory (EFT). Disentangling contributions from these different operators provides a further handle on BSM physics by both increasing the effective reach of $e^+ e^-$ colliders and distinguishing different BSM scenarios in the event of deviations from the Standard Model.

The $ZH$ production cross section provides a particularly sharp tool, as it allows a model-independent measurement of the Higgs-$Z$ coupling in Higgsstrahlung events identified solely by $Z$ recoils. The relatively large number of $ZH$ production events at proposed $e^+ e^-$ colliders is expected to give sub-percent level precision in the measurement of the $hZZ$ coupling, providing sensitivity to a range of BSM scenarios \cite{Dawson:2002wc, Englert:2013tya, Craig:2013xia, Craig:2014una, Ellis:2015sca, Drozd:2015kva, Drozd:2015rsp}. But $ZH$ production provides more than just a probe of rescalings of the $hZZ$ coupling. The measurement of angular observables in $ZH$ production provides sensitivity to a variety of tensor structures and therefore allows discrimination among a range of BSM scenarios. In this paper we investigate the potential for future circular $e^+ e^-$ colliders to discriminate between different BSM contributions to the $ZH$ production cross section through the use of angular asymmetries.

Needless to say, there is a long history of studying angular distributions of Higgs production events at $e^+ e^-$ colliders, both at LEP and for the planned ILC \cite{Hagiwara:1993sw, Gounaris:1995mx, Kilian:1996wu, GonzalezGarcia:1998wn, Hagiwara:2000tk, Han:2000mi, Barger:2003rs, Biswal:2005fh, Kile:2007ts, Dutta:2008bh, Rindani:2009pb, Contino:2013gna, Amar:2014fpa, Beneke:2014sba, Huitu:2015rha, Antusch:2015gjw, Bhattacherjee:2015xra, Modak:2013sb}. In this work we build on these previous studies by forecasting sensitivity to a complete basis of dimension-6 operators in the Higgs EFT at proposed future $e^+ e^-$ circular colliders such as CEPC and FCC-ee, accounting for realistic cut acceptance and detector effects. We believe this is the first comprehensive study of angular observables at proposed $e^+ e^-$ colliders of its kind.

Our paper is organized as follows: In Section \ref{sec:observables} we first review aspects of the Higgs EFT relevant for Higgsstrahlung, identifying a complete set of dimension-6 operators relevant for characterizing deviations in $e^+ e^- \to ZH$. We construct a complete set of angular observables in $e^+ e^- \to ZH$, closely following \cite{Beneke:2014sba}, and demonstrate their sensitivity to various dimension-6 operators at $\sqrt{s} =$ 240 GeV. In Section \ref{sec:actual} we forecast the sensitivity to these angular observables at proposed $e^+ e^-$ colliders, taking into account the effects of realistic cuts and detector effects. We then consider the sensitivity of proposed $e^+ e^-$ colliders to a range of BSM scenarios in Section \ref{sec:applications}, demonstrating the ability of angular observables to break degeneracies in rate measurements. We conclude in Section \ref{sec:conclusions} and reserve some details of the angular observables and our statistical treatment for the Appendix.

\section{Angular Observables} \label{sec:observables}

In this section we begin by reviewing aspects of the Higgs EFT relevant for the Higgsstrahlung  process, and identify a complete set of operators for characterizing various BSM contributions to  $e^+ e^- \to ZH$. We then construct a set of independent angular asymmetries suitable for identifying contributions from different operators, following \cite{Beneke:2014sba}.
\subsection{Higgs EFT for Higgsstrahlung}
\label{ssec:heft}

Given the apparent parametric separation of scales between the Higgs boson and new physics, the Higgs EFT provides a useful framework for characterizing deviations in Higgs properties from their Standard Model (SM) predictions. There are many independent operators at a given order in power-counting, including 59 dimension-6 operators. The Lagrangian out to dimension 6 prior to electroweak symmetry breaking takes the form
\begin{equation}
\mathcal{L}_{\rm eff} = \mathcal{L}_{\rm SM} + \frac{1}{\Lambda^2} \sum_{i = 1}^{59} \alpha_i \mathcal{O}_i
\end{equation}
Only a subset of these operators contribute to the $e^+ e^- \to ZH$ process, and of these many may be exchanged via field redefinitions or equations of motion.

Here we will work in terms of a minimal operator basis given in \cite{Beneke:2014sba}; for a comparable choice of basis, see \cite{Craig:2014una}.  The relevant operators defining our operator basis are given in Table \ref{tab:dim6ops}. Although there is no invariant meaning to a particular choice of basis, this basis is sufficient to characterize all dimension-6 contributions to $e^+ e^- \to ZH$ in the sense that all other operators contributing to  $e^+ e^- \to ZH$ can be re-written in terms of this operator basis plus additional operators irrelevant to $e^+ e^- \to ZH$.

\begin{table}[h]
\begin{center}{
\begin{tabular}{l l}
\hline \hline \vspace{-0.4cm} \\
$\CO_{\Phi\Box} = (\Phi^\dagger\Phi)\Box(\Phi^\dagger \Phi)$ &
$\CO_{\Phi W} = (\Phi^\dagger\Phi) W^{I}_{\mu \nu} W^{I\mu\nu}$
 \\
$\CO_{\Phi D} = (\Phi^\dagger D^\mu \Phi)^*(\Phi^\dagger D_\mu \Phi)$&
$\CO_{\Phi B} = (\Phi^\dagger \Phi)B_{\mu\nu}B^{\mu\nu}$ \\
$\CO^{(1)}_{\Phi \ell} = (\Phi^\dagger i
\overset{\leftrightarrow}{D}_\mu \Phi)(\bar\ell\gamma^\mu \ell)$
&$\CO_{\Phi W\! B} = (\Phi^\dagger \tau^I\Phi) W^{I}_{\mu\nu} B^{\mu\nu} $ \\
$\CO^{(3)}_{\Phi \ell} = (\Phi^\dagger i
\overset{\leftrightarrow}{D} \ \!\!\!^{\, I}_\mu \Phi)
(\bar\ell\gamma^\mu\tau^I \ell)$
& $\CO_{\Phi \widetilde W} = (\Phi^\dagger\Phi) \widetilde W^{I}_{\mu\nu}
W^{I\mu\nu}$    \\
$\CO_{\Phi e} = (\Phi^\dagger i
\overset{\leftrightarrow}{D}_\mu \Phi)(\bar e \gamma^\mu e)$
& $\CO_{\Phi \widetilde B} = (\Phi^\dagger\Phi) \widetilde B_{\mu\nu}B^{\mu\nu}$
\\
$\CO_{4L} = (\bar \ell \gamma_\mu \ell)(\bar \ell \gamma^\mu \ell)$ & $\CO_{\Phi \widetilde W\! B} = (\Phi^\dagger\tau ^I\Phi)
\widetilde W_{\mu\nu}^IB^{\mu\nu}$   \\ \vspace{-0.4cm} \\
\hline \hline
\end{tabular}
\caption{A complete basis of dimension-6 operators contributing to $e^+ e^- \to ZH$. Here the $\tau^I$ are the Pauli matrices.} \label{tab:dim6ops}
}\end{center}
\end{table}

After electroweak symmetry breaking, these dimension-6 operators give rise to a variety of interaction terms relevant for $e^+ e^- \to ZH$ of the form
\begin{eqnarray}
\mathcal{L_{\rm eff}} & \supset & c_{ZZ}^{(1)}  h Z_\mu Z^\mu + c_{ZZ}^{(2)} h Z_{\mu\nu} Z^{\mu\nu} + c_{Z\widetilde Z} h Z_{\mu\nu}\widetilde Z^{\mu\nu} + c_{AZ} h  Z_{\mu\nu} A^{\mu\nu} + c_{A\widetilde Z} h Z_{\mu\nu} \widetilde A^{\mu\nu} \nonumber \\
&& +  h Z_\mu \bar \ell \gamma^\mu \left( c_V + c_A\gamma_5   \right) \ell +Z_\mu \bar \ell \gamma^\mu (g_V - g_A \gamma_5)\ell - g_{\rm em} Q_\ell A_\mu \bar \ell \gamma^\mu \ell ,
\label{eq:efflag}
\end{eqnarray}
where $h$ is the real CP-even Higgs scalar, $Z_{\mu \nu}$ and $A_{\mu \nu}$ are the $Z$ boson and photon gauge field strengths, and $\tilde V^{\mu \nu} = \epsilon^{\mu \nu \alpha \beta} V_{\alpha \beta}$. Here we again use the notation of \cite{Beneke:2014sba} for clarity.

The couplings in this broken-phase effective Lagrangian may be straightforwardly expressed in terms of coefficients in the dimension-6 Higgs EFT. In this respect it is useful to consider the following linear combinations of (dimensionless) dimension-6 operator coefficients:
\begin{eqnarray}\nonumber
\alpha_{ZZ}^{(1)} &=&  \alpha_{\Phi \Box} -\frac{1}{2}\delta'_{G_F}+ \frac{1}{4}  \alpha_{\Phi D},  \\ \nonumber 
\alpha_{ZZ} &=& c_W^2 \alpha_{\Phi W} + s_W^2 \alpha_{\Phi B} + s_W c_W \alpha_{\Phi W\!B}, \\
\alpha_{Z \widetilde Z} &=& c_W^2 \alpha_{\Phi \widetilde W}+s_W^2 \alpha_{\Phi \widetilde B} + s_W c_W \alpha_{\Phi \widetilde W\!B},\\ \nonumber
\alpha_{AZ} &=& 2s_W c_W ( \alpha_{\Phi W} - \alpha_{\Phi B}) + (s_W^2-c_W^2)\alpha_{\Phi W\!B},\\ \nonumber
\alpha_{A \widetilde Z} &=& 2s_Wc_W ( \alpha_{\Phi \widetilde W}-\alpha_{\Phi \widetilde B}) + (s_W^2-c_W^2) \alpha_{\Phi \widetilde W\!B},\\ \nonumber
\alpha_{AA} &=&  s_W^2 \alpha_{\Phi W}+ c_W^2 \alpha_{\Phi B} - s_Wc_W \alpha_{\Phi W\!B}.
\label{eq:effalpha}
\end{eqnarray}
where $\delta'_{G_F} = - \alpha_{4L} + 2 \alpha_{\Phi \ell}^{(3)} $ and $\delta_{G_F} = \frac{v^2}{\Lambda^2} \delta' _{G_F}$. These are convenient linear combinations in the sense that they may be simply related to the coefficients of the broken-phase effective Lagrangian via
\begin{eqnarray} \nonumber
c_{ZZ}^{(1)}  &=& m_Z^2(\sqrt{2}G_F)^{1/2} \left(1 + \widehat\alpha_{ZZ}^{(1)}  \right),  \\ \nonumber
c_{ZZ}^{(2)} &=& (\sqrt{2}G_F)^{1/2} \widehat\alpha_{ZZ}, \\
c_{Z\widetilde Z} &=&(\sqrt{2}G_F)^{1/2} \widehat \alpha_{Z\widetilde Z}, \\ \nonumber
c_{AZ} &=&  (\sqrt{2}G_F)^{1/2} \widehat \alpha_{AZ},  \\ \nonumber
c_{A\widetilde Z} &=&(\sqrt{2}G_F)^{1/2} \widehat \alpha_{A\widetilde Z}.
\label{eq:effc}
\end{eqnarray}
where for convenience we have defined the dimensionless hatted coefficients via $\widehat \alpha_i = \alpha_i \frac{v^2}{\Lambda^2}$. Written in this way, the coefficient $c_{ZZ}^{(1)}$ decomposes into the Standard Model prediction parameterized in terms of the input parameters $m_Z, G_F, \alpha_{em}(q^2 = 0)$ plus corrections given by new physics encoded in the Higgs EFT. The remaining coefficients are zero at tree-level in the Standard Model, and so the leading contributions are taken to be those from the dimension-6 operators. Note that the CP-even coeficients are generated at one loop in the Standard Model.

Similarly, for the $hZ \ell \ell$ contact terms in Eq.~(\ref{eq:efflag}) it is useful to form the linear combinations
\begin{eqnarray} \nonumber
\widehat\alpha^V_{\Phi \ell}  &=& \widehat\alpha_{\Phi e} + \left(\widehat\alpha_{\Phi \ell}^{(1)}+ \widehat\alpha_{\Phi \ell}^{(3)}\right) \\
\widehat\alpha^A_{\Phi \ell}   &=& \widehat\alpha_{\Phi e} - \left(\widehat\alpha_{\Phi \ell}^{(1)}+ \widehat\alpha_{\Phi \ell}^{(3)}\right).
\label{eq:alphacontact}
\end{eqnarray}
which are simply related to the coefficients in Eq.~(\ref{eq:efflag}) by
\begin{eqnarray} \nonumber
c_V &=& \sqrt{2}G_F m_Z  \widehat\alpha^V_{\Phi \ell} \\
c_A &=& \sqrt{2}G_F m_Z \widehat\alpha^A_{\Phi \ell}
\label{eq:effcontact}
\end{eqnarray}
There are also shifts in the couplings between gauge bosons and leptons relative to the Standard Model prediction, given by
\begin{eqnarray} \nonumber
g_V &=& \frac{m_Z}{2} (\sqrt{2} G_F)^{1/2} \left[  \left(1- 4  s_W^2 \right) -  \delta g_V \right] \\
g_A &=& \frac{m_Z}{2} (\sqrt{2} G_F)^{1/2} \left(1 + \delta g_A\right) 
\label{eq:gVA}
\end{eqnarray}
Again, this decomposes into the tree-level Standard Model contributions plus dimension-6 contributions encoded in the linear combinations
\begin{eqnarray}
\delta g_V &=& -\widehat \alpha^V_{\Phi \ell} +\frac{\widehat \alpha_{\Phi D}}{4} +\frac{ \delta_{G_F}}{2} + \frac{4 s_W^2}{c_{2W}} \left[ \frac{\widehat\alpha_{\Phi D}}{4} + \frac{c_W}{s_W}  \widehat \alpha_{\Phi WB} + \frac{\delta_{G_F}}{2}   \right] \nonumber \\
\delta g_A &=&  - \widehat \alpha^A_{\Phi \ell} - \frac{\widehat \alpha_{\Phi D}}{4} - \frac{\delta_{G_F}}{2} 
\label{eq:dgVA}
\end{eqnarray}
This combination of direct interactions, contact terms, and shifts in the Higgs-gauge couplings accounts for the effects of the Higgs EFT on the process $e^+ e^- \to ZH$ up to dimension-6.

\subsection{Angular observables in $ZH$ production}

In general, these effective operators contribute to a shift in the cross section for $e^+ e^- \to ZH$, so that a linear combination of Wilson coefficients can be constrained to high precision by future $e^+ e^-$ colliders \cite{Craig:2014una}. However, there is additional information available in Higgsstrahlung events that allows us to constrain independent linear combinations of Wilson coefficients. This independent information can be effectively parameterized in terms of angular observables. In this paper we will work in terms of the parameterization in \cite{Beneke:2014sba}, although other definitions of angular observables are possible and in principle may prove more efficient in isolating specific Wilson coefficients.

\begin{figure}[t]
\centering
\includegraphics[width=12cm]{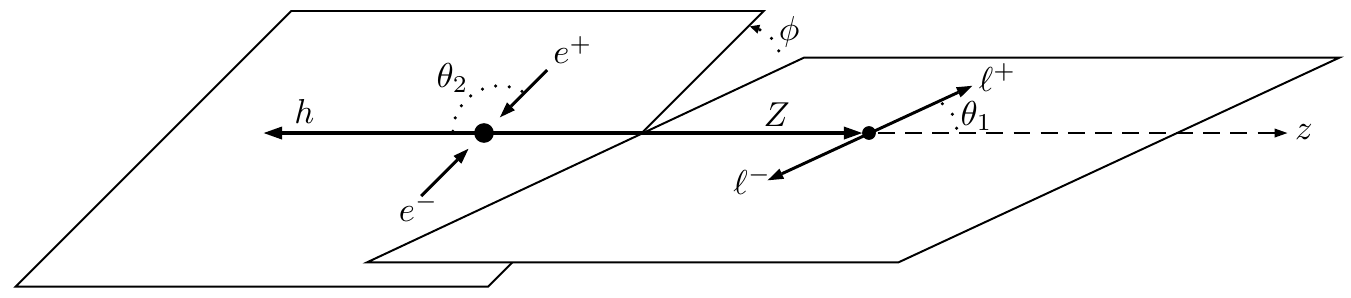}
\caption{Definition of the angles $\theta_1, \theta_2$, and $\phi$ in the $e^+ e^- \to ZH$ production process relevant for the construction of angular observables. See text for further details.}
\label{fig:angle}
\end{figure}

We define the angles $\cos \theta_1, \cos \theta_2$ and $\phi$ as follows: the $z$ direction is defined by the momentum of the on-shell $Z$ boson in the rest frame of the incoming $e^+ e^-$ pair. The $xz$ plane is the plane defined by the momentum of the outgoing $Z$ boson and its $\ell^-$ decay product. Then $\theta_1$ is the angle between the momentum of the outgoing $\ell^+$ and the $z$ axis. $\theta_2$ is the angle between the momentum of the incoming $e^+$ and the momentum of the outgoing $h$ along the $z$ axis. Finally, the angle $\phi$ corresponds to the angle in the $xy$ plane between the planes defined by the incoming $e^+ e^-$ and the outgoing $\ell^+ \ell^-$, respectively. These angles are illustrated in Fig.~\ref{fig:angle}.

In terms of these angles, we parameterize the triple differential cross section for $e^+ e^- \to Z(\to \ell^+ \ell^-)h$ as
\begin{equation}
\frac{d \sigma}{d \cos \theta_1 d \cos \theta_2 d \phi} = \frac{1}{2^{10} (2 \pi)^3} \frac{1}{\sqrt{r} \gamma_Z} \frac{\sqrt{\lambda(1,s,r)}}{s^2} \frac{1}{m_h^2} \mathcal{J}(q^2, \theta_1, \theta_2, \phi) \,,
\end{equation}
where $r = m_Z^2 / m_h^2 \approx 0.53$, $\gamma_Z = \Gamma_Z / m_h \approx 0.020$, $s = q^2 / m_h^2$, $\lambda(a,b,c) = a^2 + b^2 +c^2 - 2ab - 2 ac - 2bc$, and the function $\mathcal{J}$ contains nine independent angular structures with coefficients $J_1, \dots, J_9$ decomposed as
\begin{eqnarray} \nonumber
\mathcal{J}(q^2, \theta_1, \theta_2, \phi) = J_1 ( 1 + \cos^2 \theta_1 \cos^2 \theta_2 + \cos^2 \theta_1 + \cos^2 \theta_2) + J_2 \sin^2 \theta_1 \sin^2 \theta_2 \\
+ J_3 \cos \theta_1 \cos \theta_2 + J_4 \sin \theta_1 \sin \theta_2 \sin \phi + J_5 \sin 2 \theta_1 \sin 2 \theta_2 \sin \phi \, .
\end{eqnarray}
 The explicit form of the $J_i$ in terms of the EFT coefficients and Standard Model parameters was computed by \cite{Beneke:2014sba} and for convenience is given in Appendix \ref{app:j}. The total integrated cross section for $e^+ e^- \to ZH$ is given in terms of the $J_i$ simply by
\begin{equation}
\sigma(s) = \frac{32 \pi}{9} \frac{1}{2^{10} (2 \pi)^3} \frac{1}{\sqrt{r} \gamma_Z} \frac{\sqrt{\lambda(1,s,r)}}{s^2} \frac{1}{m_h^2} (4 J_1 + J_2) \,.  \label{eq:totalcs}
\end{equation}

It is useful to isolate various combinations of terms in the differential cross section through the following angular observables $\mathcal{A}_i$, normalized to $\sigma$:
\begin{eqnarray} \nonumber
\mathcal{A}_{\rm  \theta_1} &=& \frac{1}{\sigma}  \, \int_{-1}^1  \, d \cos \theta_1 \, {\rm sgn}(\cos (2\theta_1)) \, \frac{d \sigma }{ d\cos \theta_1}  \\
&=& 1-\frac{5}{2\sqrt{2}}+\frac{3J_1}{\sqrt{2}(4J_1+J_2)} \\
\mathcal{A}_{\rm  \phi}^{(1)} &=& \frac{1}{\sigma} \, \int_{0}^{2 \pi}  d \phi  \, {\rm sgn}(\sin \phi) \, \frac{d \sigma}{d \phi} = \frac{9  \pi}{32} \, \frac{J_4}{4J_1+J_2} \\
\mathcal{A}_{\rm  \phi}^{(2)} &=& \frac{1}{ \sigma} \, \int_{0}^{2 \pi}   d \phi  \, {\rm sgn}(\sin (2 \phi) ) \, \frac{d \sigma}{d \phi} = \frac{2}{\pi} \, \frac{J_8}{4J_1+J_2}  \\
\mathcal{A}_{\rm \phi}^{(3)} &=& \frac{1}{ \sigma} \, \int_{0}^{2 \pi} d \phi  \, {\rm sgn}(\cos \phi) \,
\frac{d \sigma}{d \phi} = \frac{9 \pi}{32} \, \frac{J_6}{4J_1+J_2}  \\
\mathcal{A}_{\rm  \phi}^{(4)} &=& \frac{1}{\sigma} \, \int_{0}^{2 \pi}   d \phi  \, {\rm sgn}(\cos (2 \phi) ) \, \frac{d \sigma}{d \phi} = \frac{2}{\pi} \,
\frac{J_{9}}{4J_1+J_2} 
\end{eqnarray}
Here ${\rm sgn}(\pm |x|) = \pm 1$. In addition to these five angular observables, it is also useful to define the forward-backward asymmetry

\begin{eqnarray}
\mathcal{A}_{c \theta_1, c \theta_2} &=&\frac{1}{\sigma } \int_{-1}^{1}
d\cos\theta_1\, {\rm sgn}(\cos \theta_1)\int_{-1}^{1}
d\cos\theta_2\, {\rm sgn}(\cos \theta_2)
\frac{d^2 \sigma}{d \cos \theta_1 d \cos \theta_2 }  \nonumber\\
&=& \frac{9}{16} \, \frac{J_{3}}{4J_1+J_2}. \label{eq:Afb}
\end{eqnarray}

Although there are nine $J_i$, only six are independent, leading to six independent angular observables corresponding to the six independent form factors in the $e^+ e^- \to ZH$ amplitude.
Each of the angular observables is sensitive to a different linear combination of coefficients in the dimension-6 Higgs EFT.

\begin{figure}[htp!]
\centering
\includegraphics[height=0.5\textwidth]{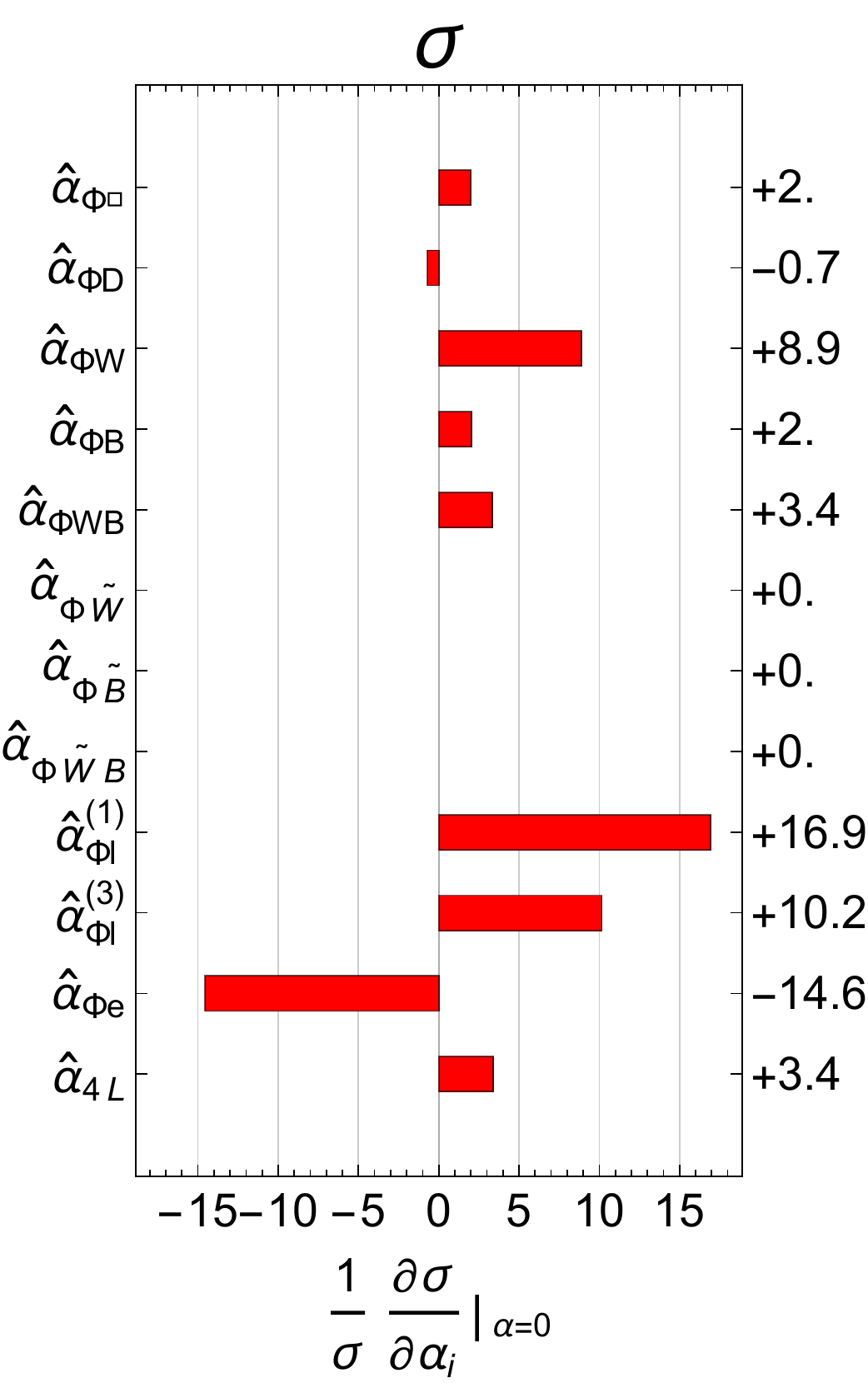}
\includegraphics[height=0.5\textwidth]{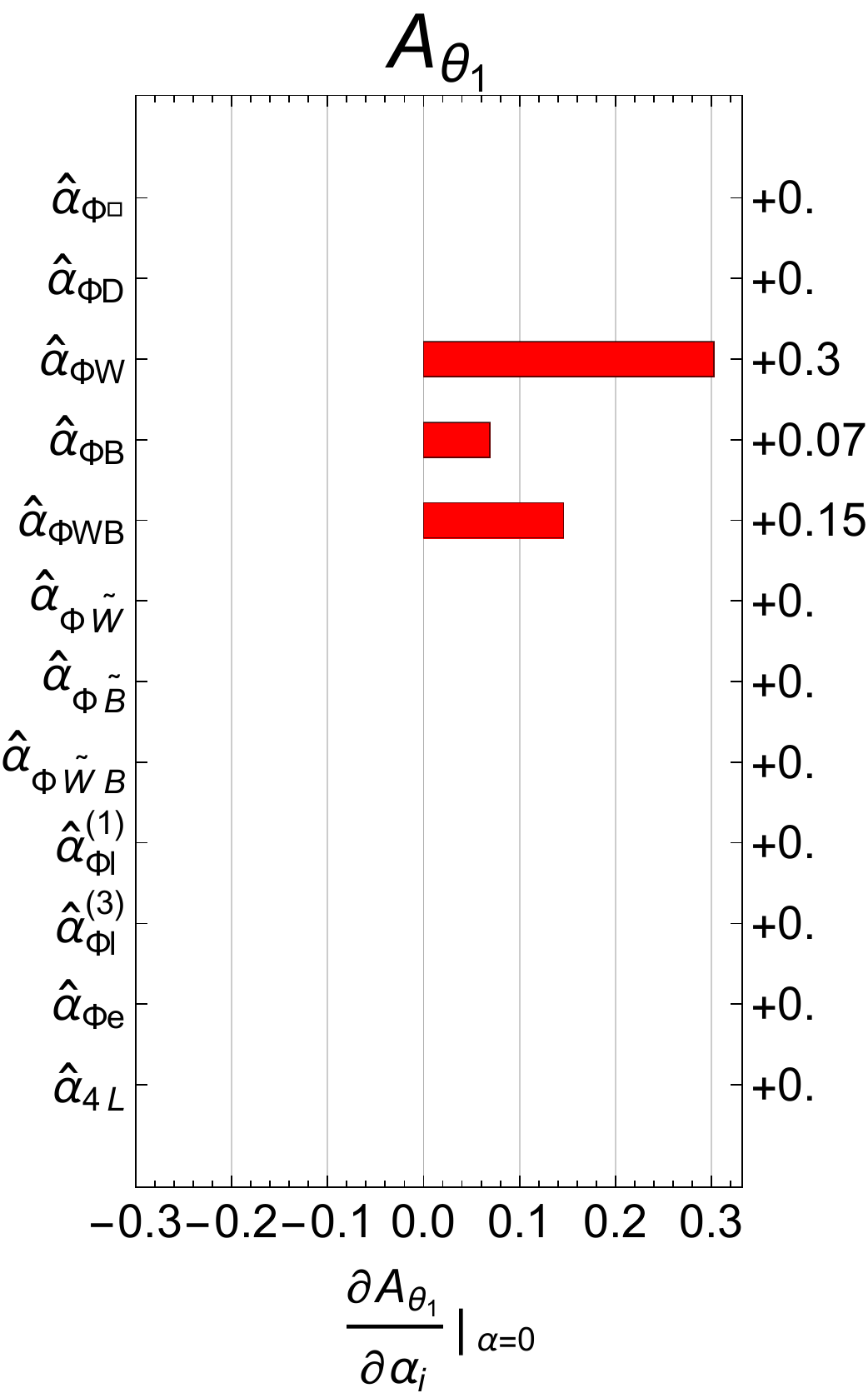}
\includegraphics[height=0.5\textwidth]{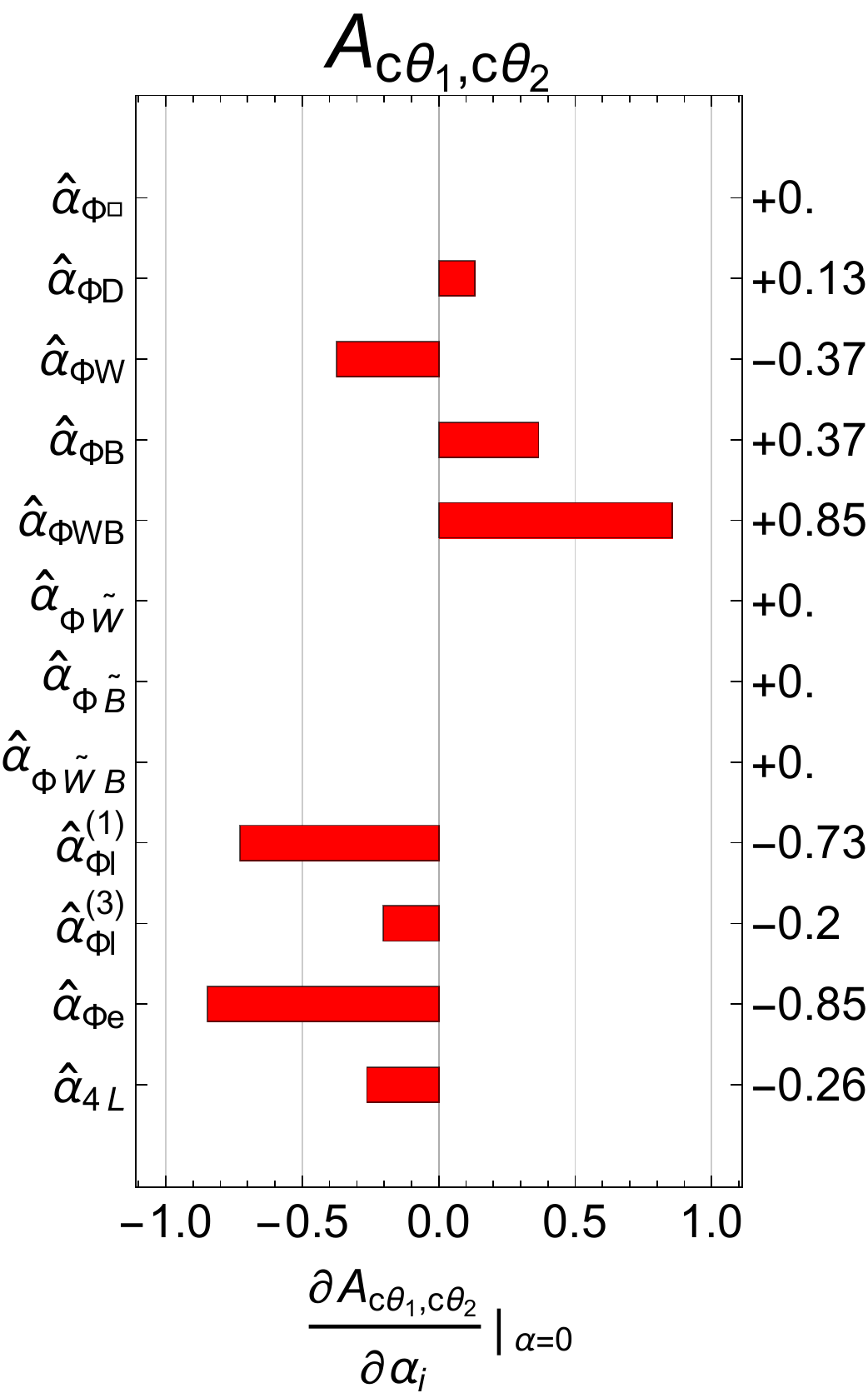}
\caption{The sensitivity of $\sigma$, $\mathcal{A}_{\rm  \theta_1}$ and $\mathcal{A}_{c \theta_1, c \theta_2}$ to different Wilson coefficients at $\sqrt{s}=240$~GeV,  corresponding to Eq.~(\ref{eq:csn})--(\ref{eq:Ac12n}).  Each row shows the value of $\frac{\partial f}{\partial \ha_i} |_{\boldsymbol{\ha}=0}$, where $f$ is the observable and $\ha_i$ is the Wilson coefficient labeled on the left side of the plot.}
\label{fig:bar1}
\end{figure}

\begin{figure}[htp!]
\centering
\includegraphics[height=0.5\textwidth]{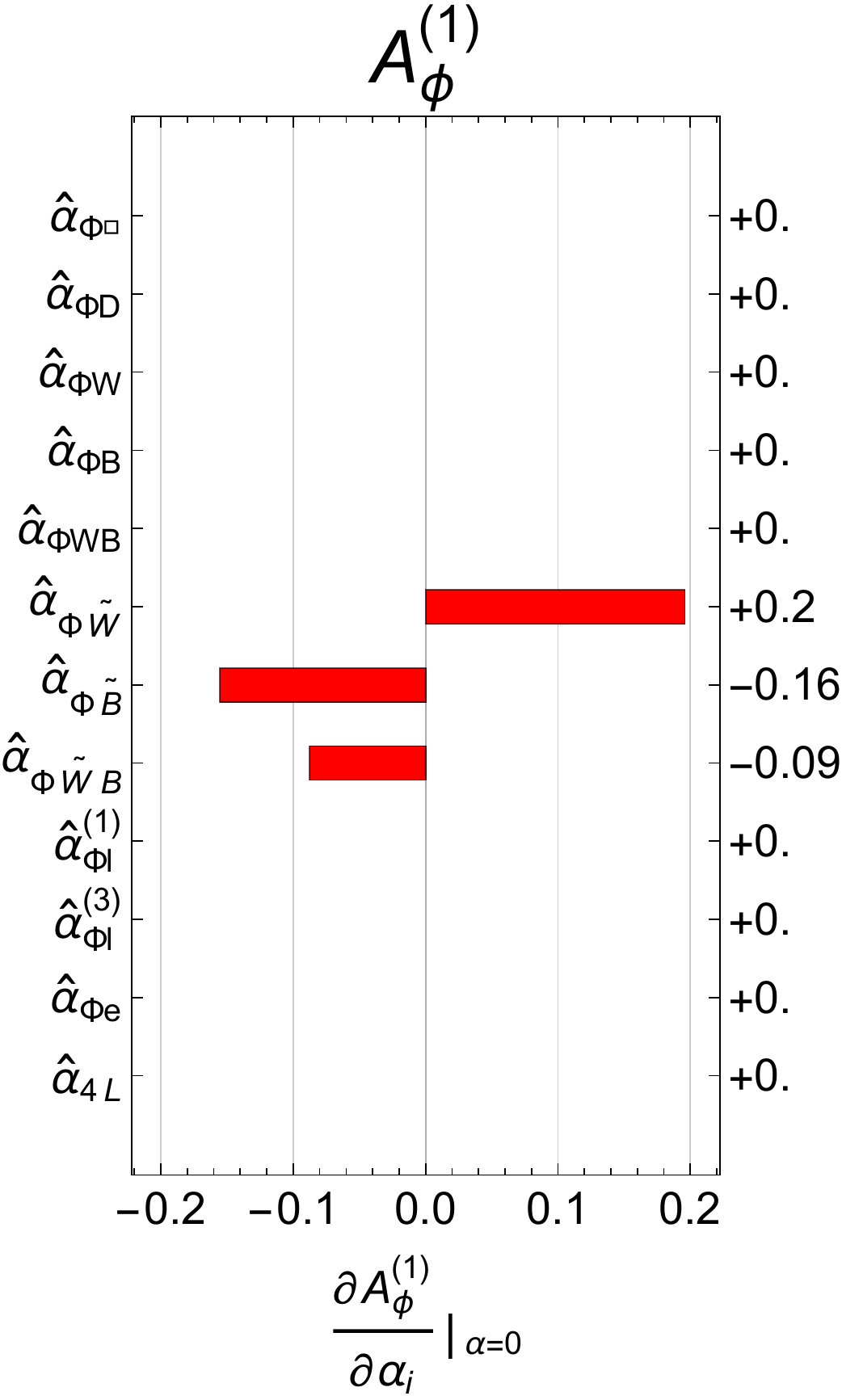}
\includegraphics[height=0.5\textwidth]{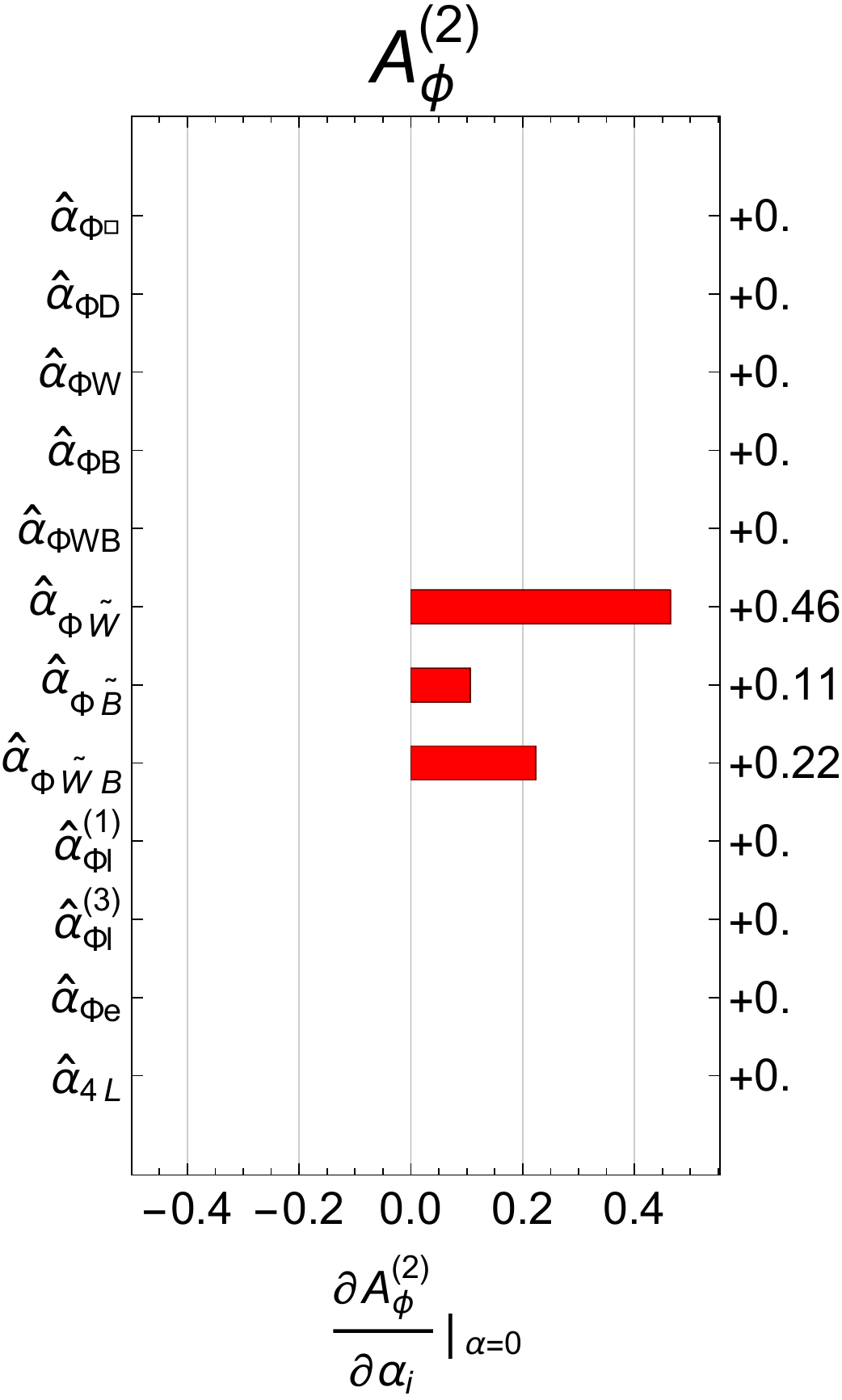}
\\ \vspace{0.3cm}
\includegraphics[height=0.5\textwidth]{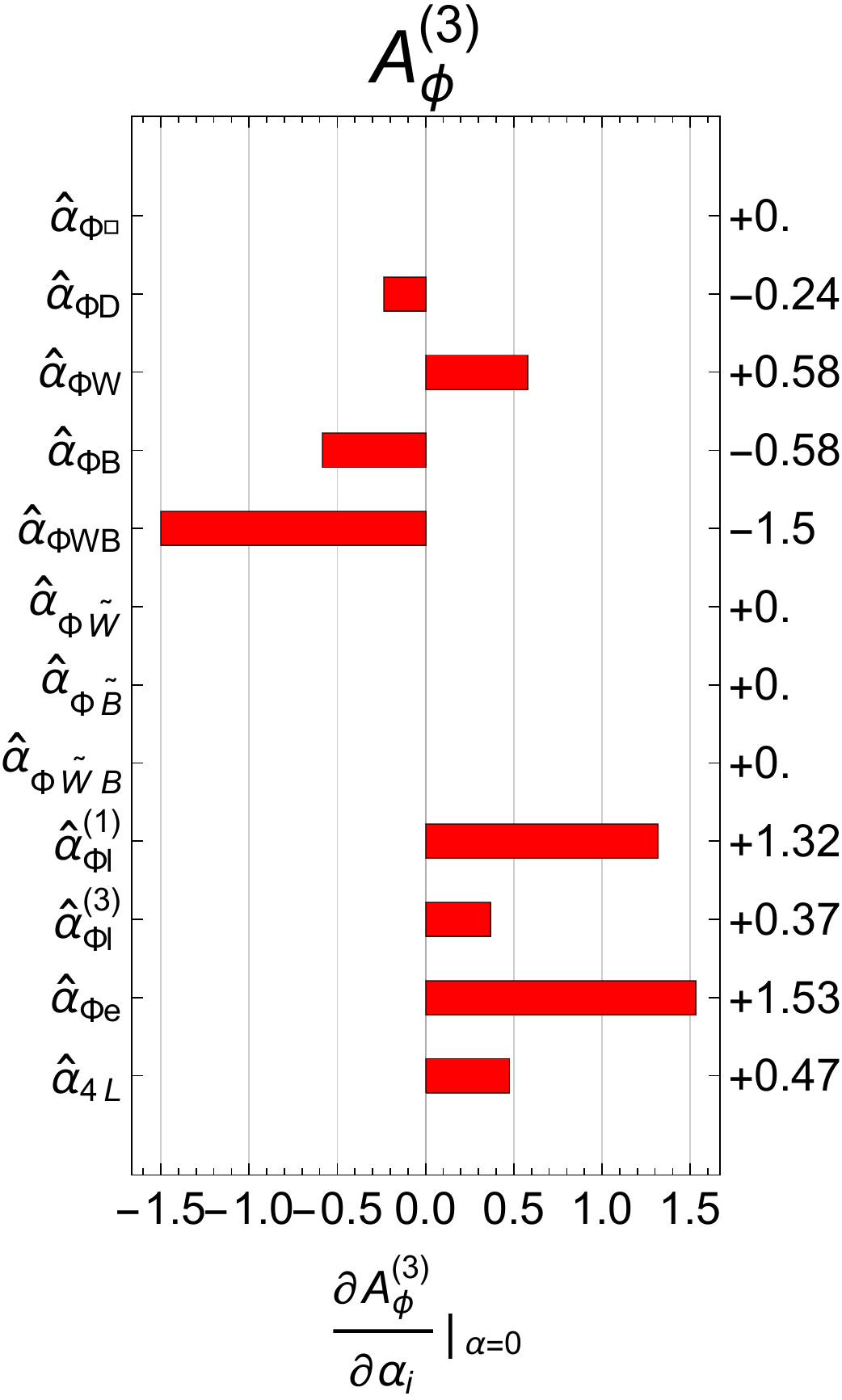}
\includegraphics[height=0.5\textwidth]{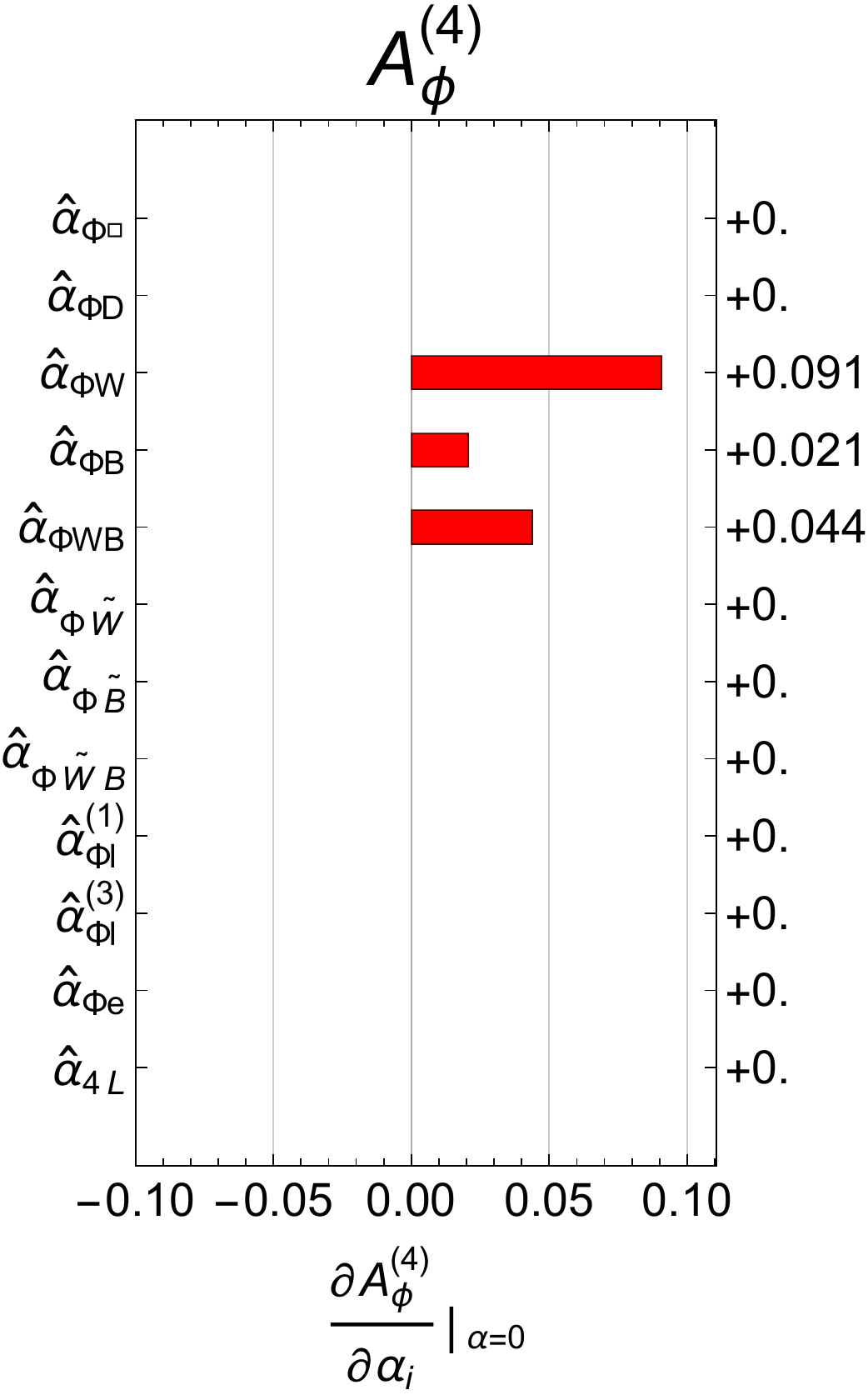}
\caption{Same as Fig.~\ref{fig:bar1} but for $\mathcal{A}_{\rm  \phi}^{(1)}$, $\mathcal{A}_{\rm  \phi}^{(2)}$, $\mathcal{A}_{\rm  \phi}^{(3)}$ and $\mathcal{A}_{\rm  \phi}^{(4)}$, which corresponds to Eq.~(\ref{eq:Aphi1n})--(\ref{eq:Aphi4n}).}
\label{fig:bar2}
\end{figure}

From Eq.~(\ref{eq:totalcs})--(\ref{eq:Afb}), it is straightforward to write down the numerical expressions for the asymmetry observables (as well as the total cross section) as a function of Wilson coefficients in the HEFT.  We choose to parameterize deviations from the Standard Model in terms of the input parameters $m_Z$, $G_F$ and $\alpha_{em}(q^2=0)$, which are very well measured.  However, we must be somewhat careful about the scale at which we evaluate couplings in physical observables.  Ideally, one should evaluate the couplings at the relevant scale of the process, $240$\,GeV; but in practice, such a procedure would face two difficulties.  First, taking count of the running effects would require a careful modification of the parameterizations in Section~\ref{ssec:heft}, which are only valid at tree level.  Second, new physics that modifies the Higgs effective Lagrangian would in principle also contribute to the running of the gauge couplings above the scale $m_h$.  A proper treatment of the running requires one to consistently account for the one-loop modification of beta functions for the electroweak gauge couplings in the presence of dimension-6 operators \cite{Jenkins:2013zja}.  On the other hand, by direct evaluating couplings at zero momentum one would omit the large logarithmic terms due to running from $m_e$ to $m_Z$ which could significantly change the results.     
For the purposes of this study, we therefore choose an intermediate scale, the $Z$-pole, at which the couplings are well-determined and the large logarithmic terms due to running from zero momentum are (mostly) accounted for.

The numerical values of the input parameters are chosen as follows.  We use the Particle Data Group (PDG) values for $m_Z$, $G_F$ \cite{Agashe:2014kda}.  We also use the PDG value $\alpha_{em}(m^2_Z)^{-1} = 127.940\pm0.014$\,, which is obtained by RG evolving $\alpha_{em}(q^2=0)$ to the $Z$-pole in the modified subtraction ($\overline{\rm MS}$) scheme.  For consistency, we use the value of $\sin^2{\theta_W}$ directly calculated from the input parameters in the $\overline{\rm MS}$ scheme \cite{Kumar:2013yoa}, $\sin^2{\theta_W}=0.23124(6)$.  This is very close to the SM best fit value from PDG, $\sin^2{\theta_W}=0.23126(5)$, an indication that SM is in very good agreement with experiments up to the $Z$-pole.
The cross section and angular observables also depend on the Higgs boson mass, $m_h$, for which we use the value from \cite{Aad:2015zhl}\footnote{The cross section also depends on the total width of the $Z$ boson, $\Gamma_Z = 2.4952\pm0.0023$\,GeV, which only contributes to the overall normalization and has no impact on our results.}.     
The numerical values of the parameters are summarized in Table~\ref{tab:input}.      
\begin{table}[t]
\centering
\begin{tabular}{|c|c|c|c|c|} \hline
 $m_Z \,[{\rm GeV}]$ & $G_F \,[{\rm GeV}^{-2}]$ & $\alpha_{em}(m^2_Z)^{-1}$ & $\sin^2{\theta_W}$  & $m_h \,[{\rm GeV}]$  \\ \hline
  $91.1876\pm0.0021$ & $1.1663787(6)\times 10^{-5}$  & $127.940\pm0.014$  & $0.23124(6)$ &  $125.09\pm0.24$       \\ \hline
\end{tabular}
\caption{The numerical values of the input parameters used in our study, taken from \cite{Agashe:2014kda}, \cite{Kumar:2013yoa} and~\cite{Aad:2015zhl}.}
\label{tab:input}
\end{table} 
The uncertainties in $\sin^2{\theta_W}$ and $m_h$ have the largest effects on the results, though they remain at the sub-percent level.  For simplicity, we will ignore the uncertainties in Table~\ref{tab:input} in our study.  

Working only to linear order in the Wilson coefficients, at $\sqrt{s}=240$~GeV the dependence of the total cross section on the hatted coefficients $\widehat{\alpha}_i$ in the unbroken-phase HEFT is
\begin{align}
\sigma{\rm [fb]} \approx&~  7.96 \, \big(1 + 2.00\, \widehat{\alpha}_{\Phi\Box} -0.70 \, \widehat{\alpha}_{\Phi D} +  8.90 \, \widehat{\alpha}_{\Phi W} + 2.03 \, \widehat{\alpha}_{\Phi B} + 3.35 \, \widehat{\alpha}_{\Phi W\! B} \nonumber\\  & \hspace{1cm} + 16.9 \, \widehat{\alpha}^{(1)}_{\Phi\, \ell} + 10.2 \, \widehat{\alpha}^{(3)}_{\Phi\, \ell} -14.6 \, \widehat{\alpha}_{\Phi\, e} + 3.39 \, \widehat{\alpha}_{4L} \big) \,, \label{eq:csn}
\end{align}
Similarly, the dependence of the angular observables on the hatted coefficients $\widehat{\alpha}_i$ in the unbroken-phase HEFT is
\begin{align}
\mathcal{A}_{\rm  \theta_1} \approx&~  -0.448 +  0.30 \, \widehat{\alpha}_{\Phi W} + 0.069 \, \widehat{\alpha}_{\Phi B} + 0.15 \, \widehat{\alpha}_{\Phi W\! B} \,, \label{eq:Atheta1n}\\
\mathcal{A}_{c \theta_1, c \theta_2} \approx&~  -0.0075 +0.13 \, \widehat{\alpha}_{\Phi D} -0.37 \, \widehat{\alpha}_{\Phi W} +0.37 \, \widehat{\alpha}_{\Phi B}  +0.85 \, \widehat{\alpha}_{\Phi W\! B} \nonumber\\
& ~~ -0.73 \, \widehat{\alpha}^{(1)}_{\Phi\, \ell} -0.20 \, \widehat{\alpha}^{(3)}_{\Phi\, \ell} -0.85 \, \widehat{\alpha}_{\Phi\, e} -0.26 \, \widehat{\alpha}_{4L} \,,   \label{eq:Ac12n}
\end{align}
and likewise
\begin{align}
\mathcal{A}_{\rm  \phi}^{(1)} \approx&~   0.20\, \widehat{\alpha}_{\Phi \widetilde W} -0.16\, \widehat{\alpha}_{\Phi \widetilde B} -0.088\, \widehat{\alpha}_{\Phi W\!  \widetilde B}  \,, \label{eq:Aphi1n}  \\
\mathcal{A}_{\rm  \phi}^{(2)} \approx&~  0.46\,  \widehat{\alpha}_{\Phi \widetilde W} + 0.11\, \widehat{\alpha}_{\Phi \widetilde B} + 0.22\, \widehat{\alpha}_{\Phi W\!  \widetilde B} \,,  \\
\mathcal{A}_{\rm  \phi}^{(3)} \approx&~ 0.0136  -0.24 \, \widehat{\alpha}_{\Phi D} +  0.58 \, \widehat{\alpha}_{\Phi W} -0.58 \, \widehat{\alpha}_{\Phi B}  -1.50 \, \widehat{\alpha}_{\Phi W\! B} \nonumber\\
& ~~ + 1.32 \, \widehat{\alpha}^{(1)}_{\Phi\, \ell} + 0.37 \, \widehat{\alpha}^{(3)}_{\Phi\, \ell} +1.53 \, \widehat{\alpha}_{\Phi\, e} + 0.47 \, \widehat{\alpha}_{4L} \,,   \\
\mathcal{A}_{\rm  \phi}^{(4)} \approx&~ 0.0959 +  0.091 \, \widehat{\alpha}_{\Phi W} + 0.021 \, \widehat{\alpha}_{\Phi B} + 0.044 \, \widehat{\alpha}_{\Phi W\! B} \,.  \label{eq:Aphi4n}  
\end{align}
We present Eq.~(\ref{eq:csn})--(\ref{eq:Aphi4n}) graphically in Figs.~\ref{fig:bar1}~\&~\ref{fig:bar2}. The inclusive cross section $\sigma$ is unsurprisingly sensitive to all CP-even operators, with particular sensitivity to operators that shift the couplings between gauge bosons and leptons, as well as those that generate new $h Z \ell \ell$ contact terms. The asymmetry variables $\mathcal{A}_{\theta_1}$ and $\mathcal{A}_{\rm \phi}^{(4)}$ provide independent sensitivity to the operators $\CO_{\Phi W}, \CO_{\Phi B}, \CO_{WB}$, which are also the operators in this basis that are constrained by measurements of $h \to \gamma \gamma$. The forward-backward asymmetry $\mathcal{A}_{c \theta_1, c \theta_2}$ and the angular asymmetry $\mathcal{A}_{\rm \phi}^{(3)}$ are sensitive to independent linear combinations of the CP-even operators (excepting $\CO_{\Phi \Box}$, whose contribution has been eliminated by construction in the angular asymmetries). Finally, the asymmetries $\mathcal{A}_{\rm  \phi}^{(1)}$ and $\mathcal{A}_{\rm  \phi}^{(2)}$ are sensitive to independent linear combinations of CP-odd operators. 

\begin{figure}[htp!]
\centering
\includegraphics[height=0.5\textwidth]{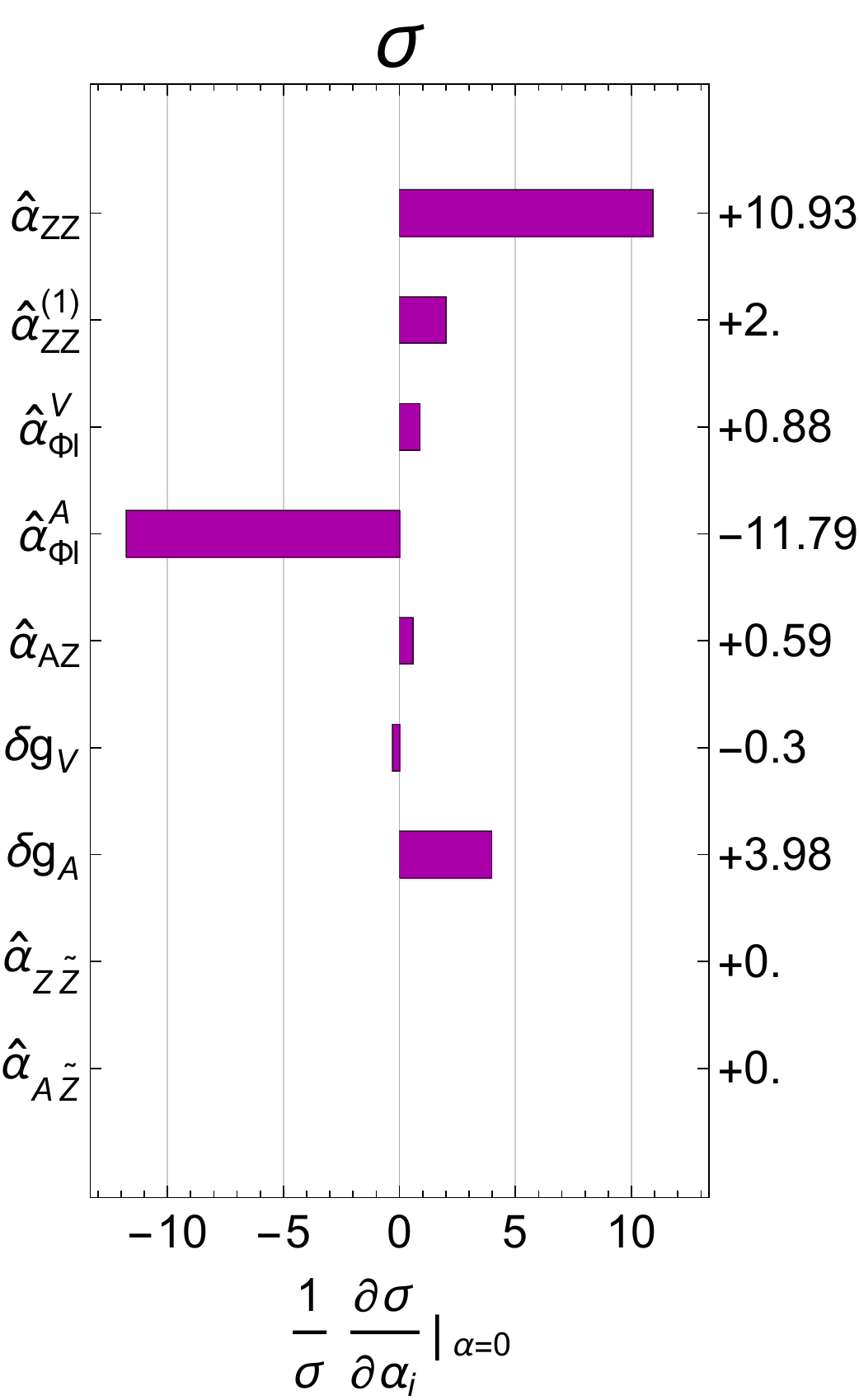}
\includegraphics[height=0.5\textwidth]{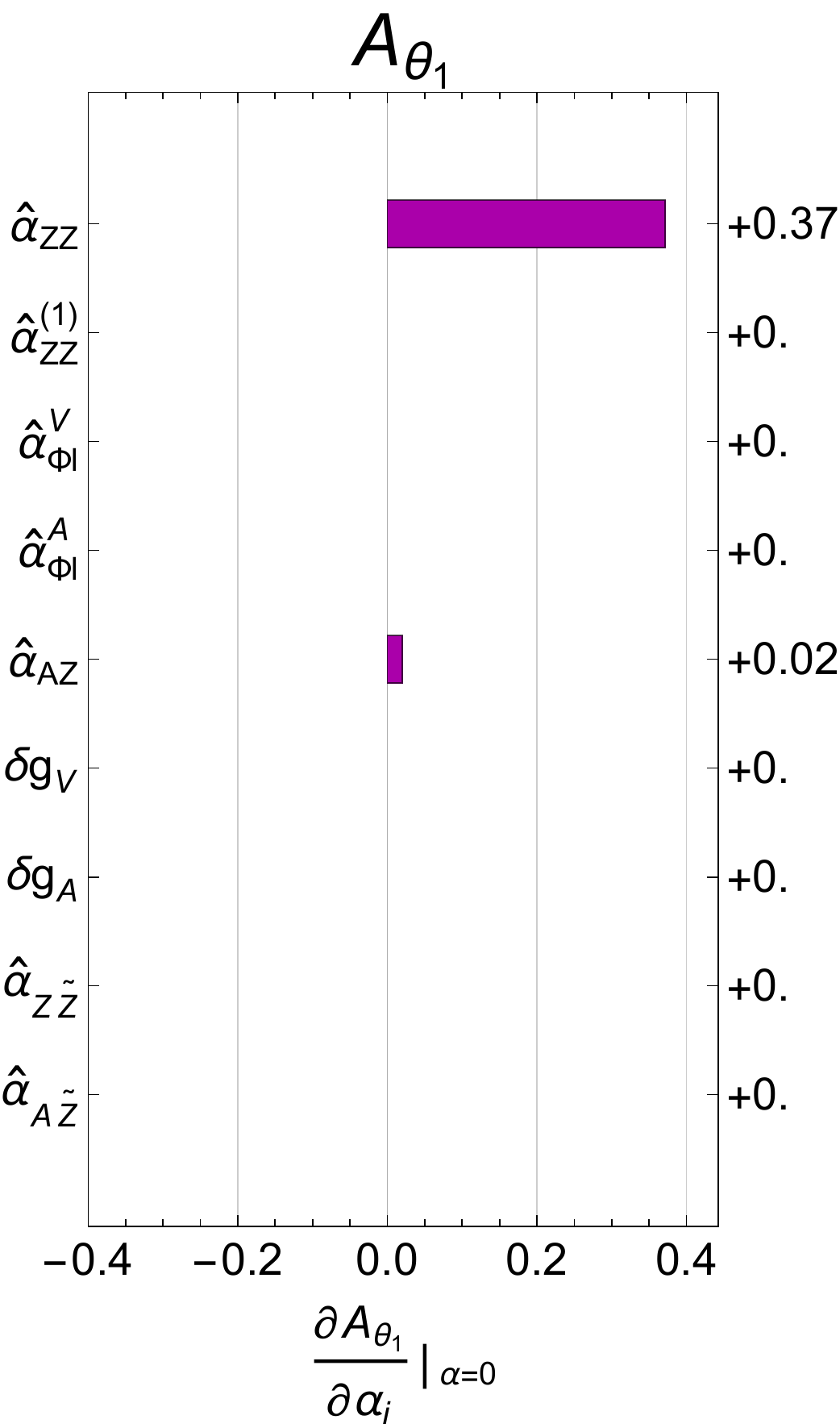}
\includegraphics[height=0.5\textwidth]{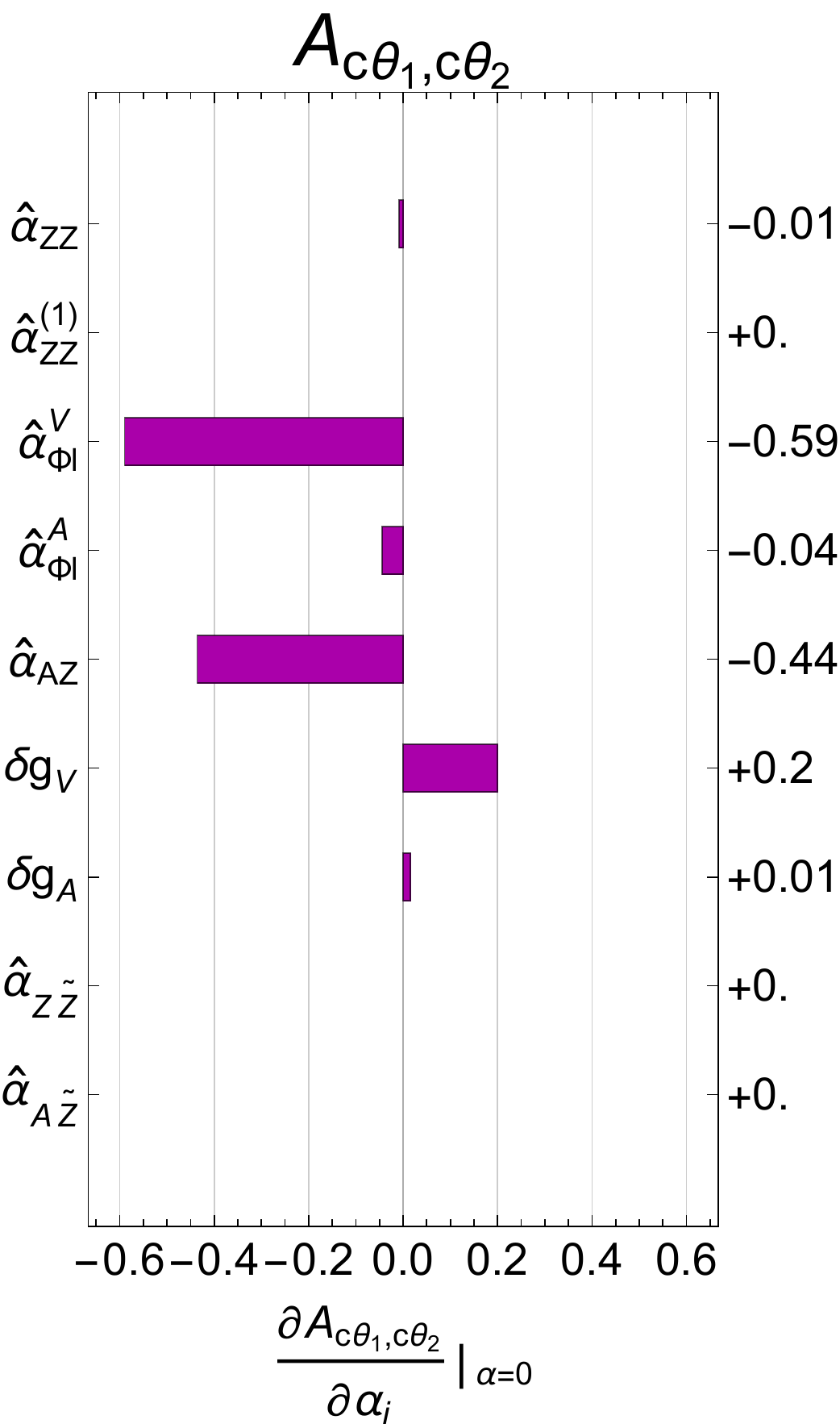}
\caption{Same as Fig.~\ref{fig:bar1} but for the sensitivity of $\sigma$, $\mathcal{A}_{\rm  \theta_1}$ and $\mathcal{A}_{c \theta_1, c \theta_2}$ to the coefficients in Eq.~(\ref{eq:wclist}) at $\sqrt{s}=240$~GeV, corresponding to Eq.~(\ref{eq:acsn})--(\ref{eq:aAc12n}).}
\label{fig:abar1}
\end{figure}
\begin{figure}[htp!]
\centering
\includegraphics[height=0.5\textwidth]{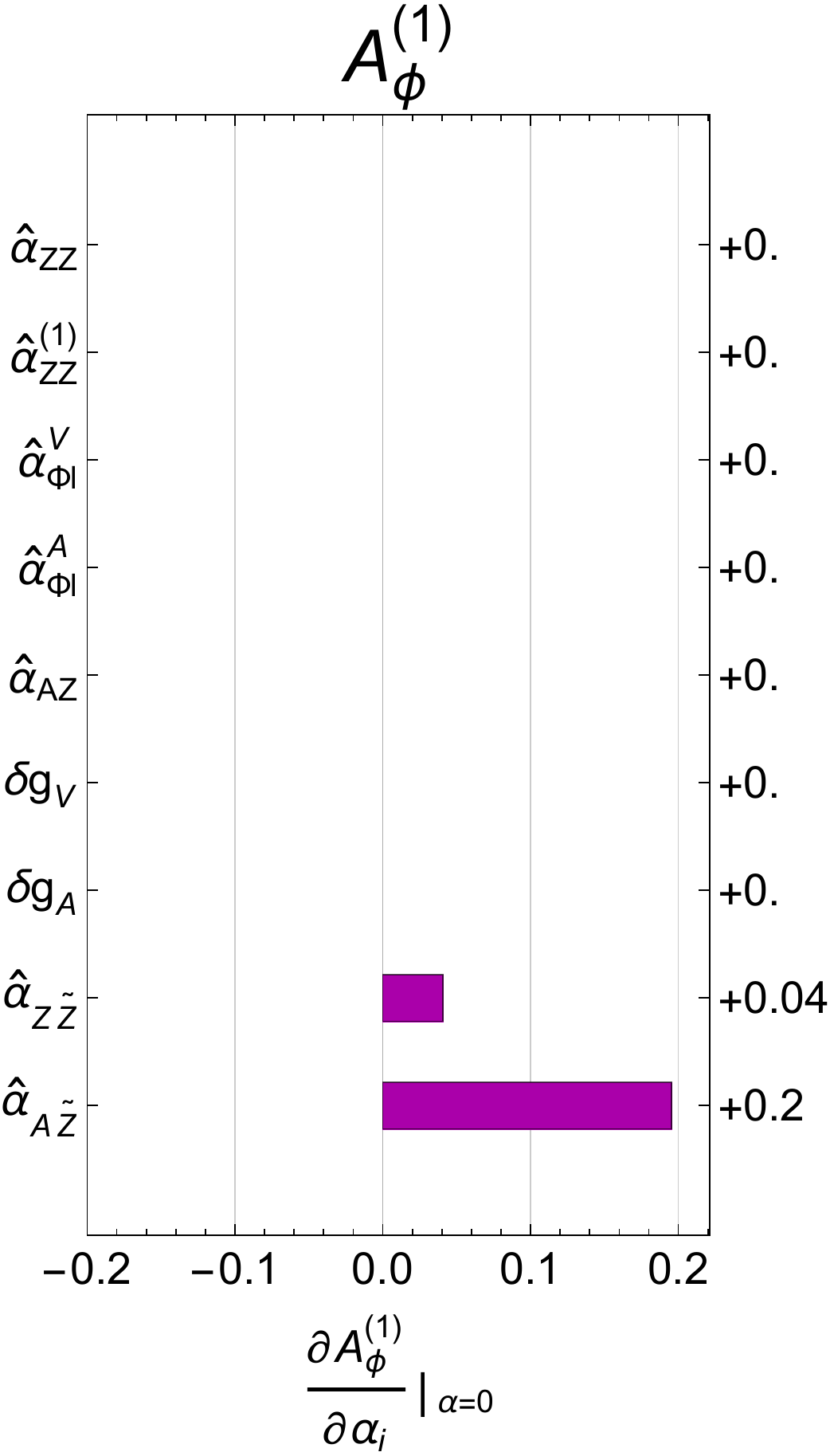}
\includegraphics[height=0.5\textwidth]{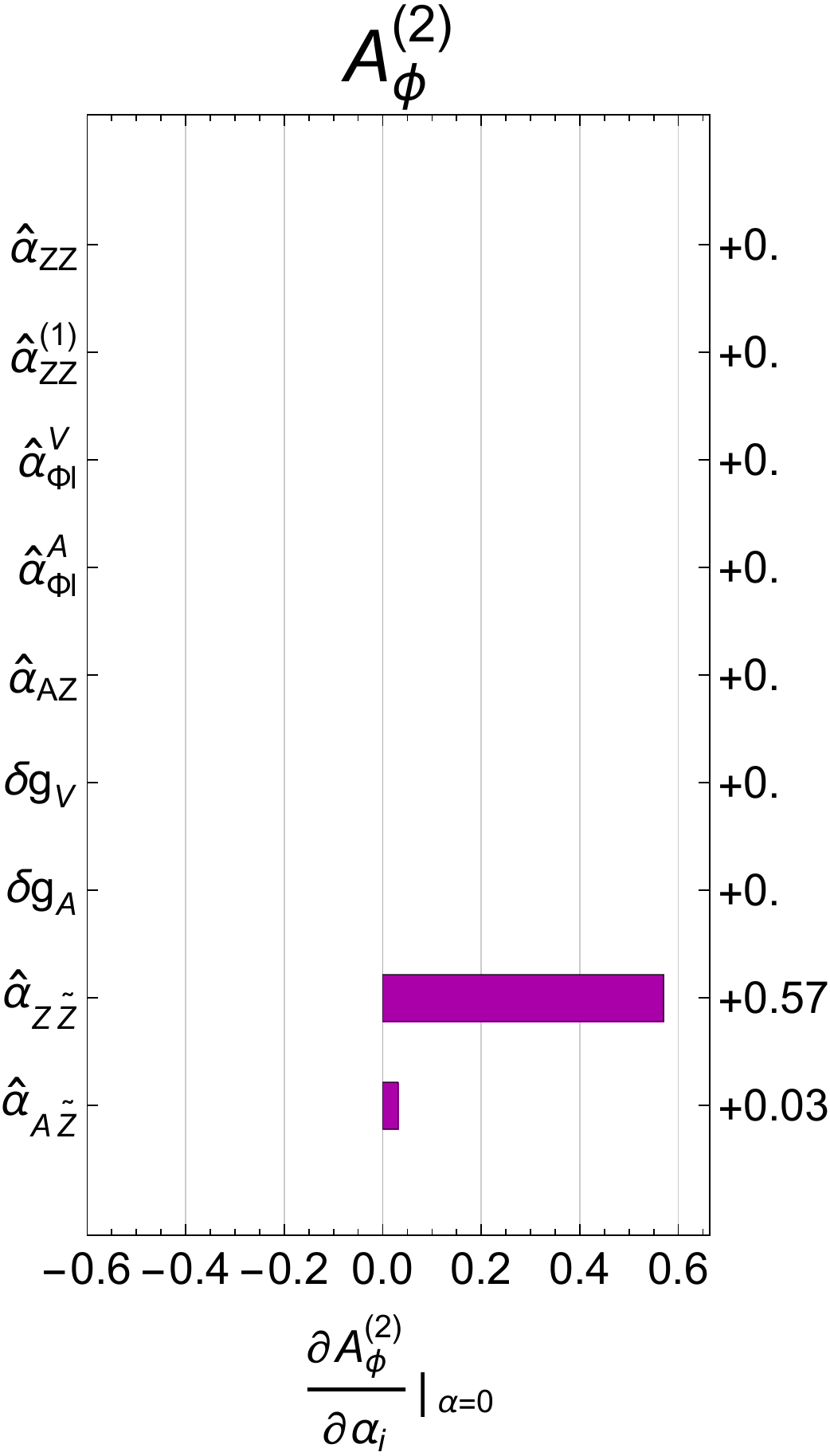}
\\ \vspace{0.3cm}
\includegraphics[height=0.5\textwidth]{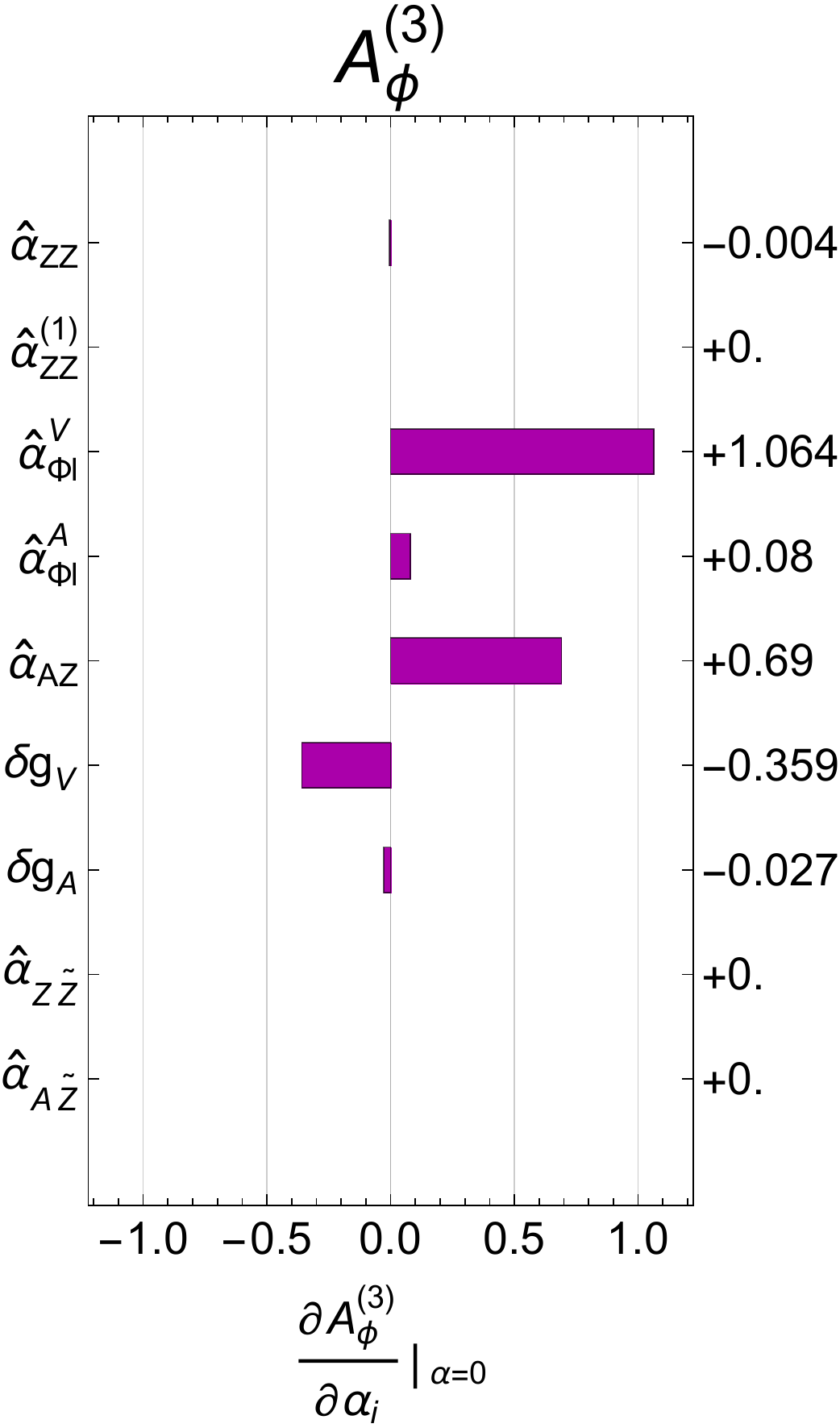}
\includegraphics[height=0.5\textwidth]{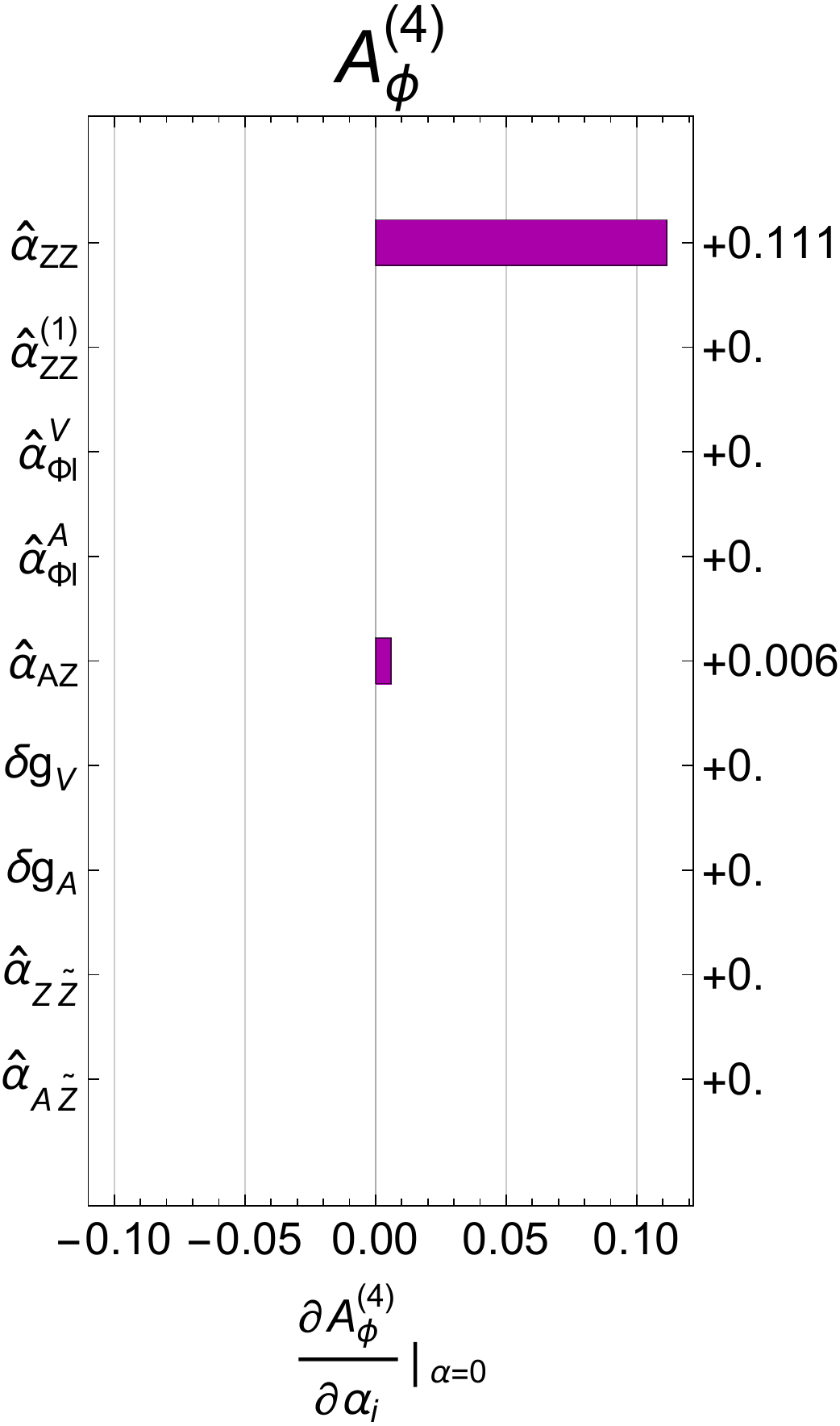}
\caption{Same as Fig.~\ref{fig:abar1} but for $\mathcal{A}_{\rm  \phi}^{(1)}$, $\mathcal{A}_{\rm  \phi}^{(2)}$, $\mathcal{A}_{\rm  \phi}^{(3)}$ and $\mathcal{A}_{\rm  \phi}^{(4)}$, which corresponds to Eq.~(\ref{eq:aAphi1n})--(\ref{eq:aAphi4n}).}
\label{fig:abar2}
\end{figure}

It is also helpful to connect the cross section and angular observables directly to the broken-phase effective Lagrangian in Eq.~(\ref{eq:efflag}).  We choose the basis with the following 9 independent coefficients:
\begin{equation}
\ha_{ZZ} \,, ~~~ \ha^{(1)}_{ZZ} \,, ~~~  \ha^V_{\Phi \ell} \,, ~~~ \ha^A_{\Phi \ell} \,, ~~~ \ha_{AZ} \,, ~~~ \delta g_V \,, ~~~ \delta g_A \,, ~~~  \ha_{Z\widetilde{Z}} \,, ~~~  \ha_{A\widetilde{Z}} \,,  \label{eq:wclist}
\end{equation}
which are related to the couplings in Eq.~(\ref{eq:efflag}) by Eqs.~(\ref{eq:effc}), (\ref{eq:effcontact}) and (\ref{eq:gVA}).  In terms of the coefficients in Eq.~(\ref{eq:wclist}), at $\sqrt{s}=240$~GeV the cross section scales as
\begin{align}
\sigma{\rm [fb]} \approx&~  7.96\,\big(1 + 10.9\, \ha_{ZZ} + 2.00\, \ha^{(1)}_{ZZ} + 0.88\, \ha^V_{\Phi \ell} -11.8\, \ha^A_{\Phi \ell}  \nonumber\\ 
&\hspace{1cm}  +  0.59\, \ha_{AZ} -0.30\, \delta g_V + 3.98\, \delta g_A \big) \,, \label{eq:acsn}
\end{align}
while the angular observables scale as 
\begin{align}
\mathcal{A}_{\rm  \theta_1} \approx&~  -0.448  + 0.37\, \ha_{ZZ} +  0.020\, \ha_{AZ} \,, \label{eq:aAtheta1n} \\
\mathcal{A}_{c \theta_1, c \theta_2} \approx&~   -0.0075 -0.0088\, \ha_{ZZ} -0.59 \, \ha^V_{\Phi \ell} -0.044 \, \ha^A_{\Phi \ell} \nonumber\\
&~~ -0.44\, \ha_{AZ}  +0.20\, \delta g_V + 0.015 \, \delta g_A   \,, \label{eq:aAc12n}
\end{align}
and
\begin{align}
\mathcal{A}_{\rm  \phi}^{(1)} \approx&~ 0.040 \, \ha_{Z\widetilde{Z}} + 0.20 \, \ha_{A\widetilde{Z}} \,,  \label{eq:aAphi1n} \\
\mathcal{A}_{\rm  \phi}^{(2)} \approx&~  0.57 \, \ha_{Z\widetilde{Z}} + 0.031 \, \ha_{A\widetilde{Z}}  \,, \\
\mathcal{A}_{\rm  \phi}^{(3)} \approx&~   0.0136 -0.0041\, \ha_{ZZ} + 1.06\, \ha^V_{\Phi \ell} + 0.080\, \ha^A_{\Phi \ell} \nonumber\\
&~~ + 0.69\, \ha_{AZ} -0.36 \, \delta g_V -0.027 \, \delta g_A  \,, \\
\mathcal{A}_{\rm  \phi}^{(4)} \approx&~  0.0959 + 0.11\, \ha_{ZZ} +  0.0060\, \ha_{AZ} \,,  \label{eq:aAphi4n}
\end{align}
These results are represented graphically in Figs.~\ref{fig:abar1}~\&~\ref{fig:abar2}. The information is equivalent to that in Figs.~\ref{fig:bar1}~\&~\ref{fig:bar2}, but provides a stronger indication of the relevant physics in the broken-symmetry phase. The total cross section is most sensitive to operators of the form $h Z_\mu Z^\mu$, $h Z_{\mu \nu} Z^{\mu \nu}$, the axial contact term $h Z_\mu \bar \ell \gamma^\mu \gamma_5 \ell$, and shifts in the axial $Z$-lepton coupling $Z_\mu \bar \ell \gamma^\mu \gamma_5 \ell$.  The asymmetry variables $\mathcal{A}_{\theta_1}$ and $\mathcal{A}_{\rm \phi}^{(4)}$ are particularly sensitive to $h Z_{\mu \nu} Z^{\mu \nu}$. The forward-backward asymmetry $\mathcal{A}_{c \theta_1, c \theta_2}$ and the angular asymmetry $\mathcal{A}_{\rm \phi}^{(3)}$ are particularly sensitive to $h Z_{\mu \nu} A^{\mu \nu}$, the vector contact term $h Z_\mu \bar \ell \gamma^\mu \ell$, and shifts in the vector $Z$-lepton coupling $Z_\mu \bar \ell \gamma^\mu \ell$. The asymmetries $\mathcal{A}_{\rm  \phi}^{(1)}$ and $\mathcal{A}_{\rm  \phi}^{(2)}$ are sensitive to complementary CP-violating terms $h A_{\mu \nu} \tilde Z^{\mu \nu}$ and $h Z_{\mu \nu} \tilde Z^{\mu \nu}$, respectively.

It is worth pointing out that the choice of scales at which the couplings are evaluated has a significant impact on the central values of asymmetry observables.  In particular, $\mathcal{A}^{(1)}_\phi$, $\mathcal{A}^{(3)}_\phi$, and $\mathcal{A}_{c\theta_1,c\theta_2}$ are proportional to the coupling combination $\left(\frac{g_V g_A}{g^2_V+g^2_A}\right)^2$, which is quite sensitive to the value of $\sin^2{\theta_W}$.  Evaluating the couplings at $\sim240$\,GeV instead of the $Z$-pole, the central values of asymmetry variables varies by $\mathcal{O}(10\%)$.  Although our choice of scales is adequate for forecasting sensitivity to asymmetry variables, a careful treatment of the Standard Model prediction for these asymmetry observables will ultimately be required in order to extract useful information from angular observables at future $e^+ e^-$ colliders.


\section{Angular Observables at CEPC and FCC-ee} \label{sec:actual}

Having defined the set of angular variables relevant for probing the Higgs EFT in $e^+ e^- \to ZH$, we now develop projections for the sensitivity attainable at various proposed Higgs factories. In particular, we study the reach in angular observables at two proposed future $e^+ e^-$ colliders: the Circular Electron-Positron Collider (CEPC) and the $e^+ e^-$ mode of the CERN Future Circular Collider (FCC-ee). Both of these colliders are designed to produce large numbers of $e^+ e^- \rightarrow Z H$ events at the center-of-mass energy $\sqrt{s} =$ 240 GeV. With a proposed luminosity of $2 \times 10^{34} \,\,\rm{cm^{-2}s^{-1}}$ per Interaction Point (IP), the integrated luminosity at CEPC will be $5\, {\rm ab}^{-1}$ over a running time of 10 years with 2 IPs \cite{CEPCPreCDR,CEPCPreCDRvolumn2}. The machine parameters of FCC-ee \cite{FCCee} project that its luminosity can reach $6 \times 10^{34} \,\,\rm{cm^{-2}s^{-1}}$ at $\sqrt{s} =$ 240 GeV, which is  three times that of CEPC. In addition, there is a factor of 2 increase in luminosity on account of the projected 4 IPs at FCC-ee, bringing the total FCC-ee luminosity to six times that of CEPC. Considering the same running time of 10 years, we therefore take the integrated luminosity at FCC-ee to be $30\, {\rm ab}^{-1}$ for the purpose of our projections.

\subsection{Expected precision and statistical uncertainty}

\begin{table}[h]
\centering
\begin{tabular}{|c|c||c|c|c|} \hline
& & \multicolumn{3}{|c|}{Precision $\sigma_A$ } \\  \cline{3-5}
observable & SM expectation   &  $5\,{\rm ab}^{-1}$  &  $30\,{\rm ab}^{-1}$ & \multirow{2}{*}{Full Stat.}\\ \cline {3-4}
& & CEPC & FCC-ee & \\ \hline\hline
$\mathcal{A}_{\theta_1}$                     &  -0.448    &  0.0060  & 0.0025   &  0.00078   \\ \hline
$\mathcal{A}^{(1)}_\phi$                      &  0            &  0.0067  &  0.0027  & 0.00087  \\ \hline
$\mathcal{A}^{(2)}_\phi$                      &  0            &  0.0067  &  0.0027  & 0.00087  \\ \hline
$\mathcal{A}^{(3)}_\phi$                      &  0.0136   &  0.0067 &  0.0027  & 0.00087  \\ \hline
$\mathcal{A}^{(4)}_\phi$                      &  0.0959   &  0.0067  &  0.0027  &  0.00086 \\ \hline
$\mathcal{A}_{c\theta_1,c\theta_2}$   &  -0.0075  & 0.0067  &  0.0027  &  0.00087 \\ \hline
\end{tabular}
\caption{The SM expectation at $\sqrt{s}=240$~GeV for the asymmetry observables and the standard deviation ($\sigma_A$) calculated from Eq.~(\ref{eq:sigA2}) for different sample sizes.  We consider the process with $Z\to \mu^+\mu^- / e^+e^-$ and $H\to b\bar{b}$, which is almost entirely background-free.  According to Section~3.3.3.1 in the CEPC pre-CDR~\cite{CEPCPreCDR}, the number of events after basic cuts is 22100 for $5\,{\rm ab}^{-1}$.  We use this number here and also scale it up with luminosity for $30\,{\rm ab}^{-1}$ and the full statistics scenario detailed in text. 
}
\label{tab:Ath}
\end{table}

In Table~\ref{tab:Ath} we list the theoretical expectations for all the relevant asymmetry observables assuming only Standard Model contributions, as well as the 1$\sigma$ errors ($\sigma_A$) calculated from Eq.~(\ref{eq:sigA2}) for various integrated luminosity benchmarks. We consider only the process with $Z\to \mu^+\mu^- / e^+e^-$ and $H\to b\bar{b}$, which is almost entirely background-free.  According to Section~3.3.3.1 in the pre-CDR of CEPC \cite{CEPCPreCDR}, the number of events after basic cuts in both $\mu^+\mu^-$ and $e^+e^-$ channels is $11067+11033=22100$ for $5\,{\rm ab}^{-1}$. We assume for simplicity that FCC-ee will conduct a very similar study on this channel, and consequently scale the statistics up directly to 30~ab$^{-1}$ for FCC-ee. 

We emphasize that our study is very conservative by focusing only on the $Z\to \mm/\ee$ decays,  which comprise only about 7\% of all $Z$ decays. A more comprehensive study could employ other visible $Z$ boson decays to improve signal statistics, albeit at the cost of requiring a detailed background analysis. If all visible decays of the $Z$ boson --  about 80\% of the total branching ratio --  and additional decay channels of the Higgs beyond $b\bar{b}$ can be included, it is in principle possible to gain a factor of ten improvement in the statistics for our analysis. Therefore, although we do not perform such an analysis here, to demonstrate the full diagnostic power of asymmetry observables we consider a third benchmark with ten times larger statistics than the FCC-ee case with only the dilepton channel.\footnote{This corresponds to $300~\abi$ integrated luminosity of the dilepton channel alone.} We refer to this as the ``Full Statistics'' (FS) benchmark with an eye towards the maximum sensitivity obtainable by using additional $Z$ and $H$ decay products.

\subsection{Simulation procedure and detector effects}
\label{ssec:Exp}

At $\sqrt{s} =240$~GeV, the dominant Higgs production process will be $\ee \rightarrow Z H$. As discussed above, we choose the low-background, precisely-measurable process $\ee\to Z H\to \mm\,b\bar{b}$ to demonstrate the diagnostic power of angular observables and their complementarity to inclusive observables such as the rate measurement.

\begin{table}[t]
\centering
\begin{tabular}{|c||c|c|c|c|c|c|}
\hline

Selection & $\mathcal{A}_{\theta_1}$ & $\mathcal{A}^{(1)}_\phi$ & $\mathcal{A}^{(2)}_\phi$ & $\mathcal{A}^{(3)}_\phi$ & $\mathcal{A}^{(4)}_\phi$ & $\mathcal{A}_{c\theta_1,c\theta_2}$ \\
\hline\hline
Initial& -0.46 & 0.0013 & 0.00076 & 0.013 & 0.093 & -0.0054 \\
\hline
$10^\circ<\theta_\mu< 170^\circ$& -0.46 & 0.0013 & 0.00063 & 0.012 & 0.057 & -0.0053 \\
\hline
$10~\gev< p_T(\mm) <90~\gev$& -0.46 & 0.0011 & 0.00070 & 0.012 & 0.058 & -0.0054 \\
\hline
$81~\gev<m_{\mm}<101~\gev$& -0.46 & 0.0009 & 0.00055 & 0.012 & 0.058 & -0.0056 \\
\hline
$120~\gev <m_{\rm recoil}<150~\gev$& -0.46 & 0.0009 & 0.00055 & 0.012 & 0.058 & -0.0056 \\
\hline
\end{tabular}
\caption[]{The summary of asymmetry observables for SM after selection cuts at $\sqrt{s} = $ 240 GeV using simulation sample using {\tt Madgraph5}.
The production process is $e^+ e^- \rightarrow H Z$, with $H \rightarrow b \bar{b}$ and $Z \rightarrow \mu^+\mu^-$.
}
\label{tab:cuts}
\end{table}

To understand the detailed effects of selection cuts on BSM contributions, we perform a numerical study on the signals using {\tt Madgraph5}~\cite{Alwall:2014hca} with dimension-six operator model file generated via {\tt FeynRules}~\cite{Christensen:2008py}.  The signal process is very clean.
For our study, we employ the event selection of CEPC preCDR analysis~\cite{CEPCPreCDR} for this channel, including lepton angular acceptance of $10^\circ<\theta_\mu< 170^\circ$, lepton pair $p_T$ of $10~\gev< p_T(\mm) <90~\gev$, lepton pair on-shell condition of $81~\gev<m_{\mm}<101~\gev$, recoil mass of $120~\gev <m_{\rm recoil}<150~\gev$ and b-tagging.\footnote{We do not include the cut on Higgs polar angle $|\cos(\theta_H)|<0.8$ for the $Z\to\mm$ sample, which is used in the CEPC study inherited from $Z\to\ee$ sample to suppress the Bhabha background. This cut is not necessary for the $Z\to\mm$ sample from our perspective.} After applying these cuts, the background is negligible and the dominant uncertainty is due to signal statistics.

Table \ref{tab:cuts} shows the cut flow for the 6 angular observables at each stage of event selection. As shown in the line 3 of this table, the acceptance of the polar angle can greatly change some anglular observables. This suggests, among other things, that the power of angular observables might be improved at future $e^+ e^-$ colliders by enlarging the detector acceptance for charged leptons beyond current projections.

\begin{figure}[]
\centering
\subfigure{
\includegraphics[width=0.485\textwidth]{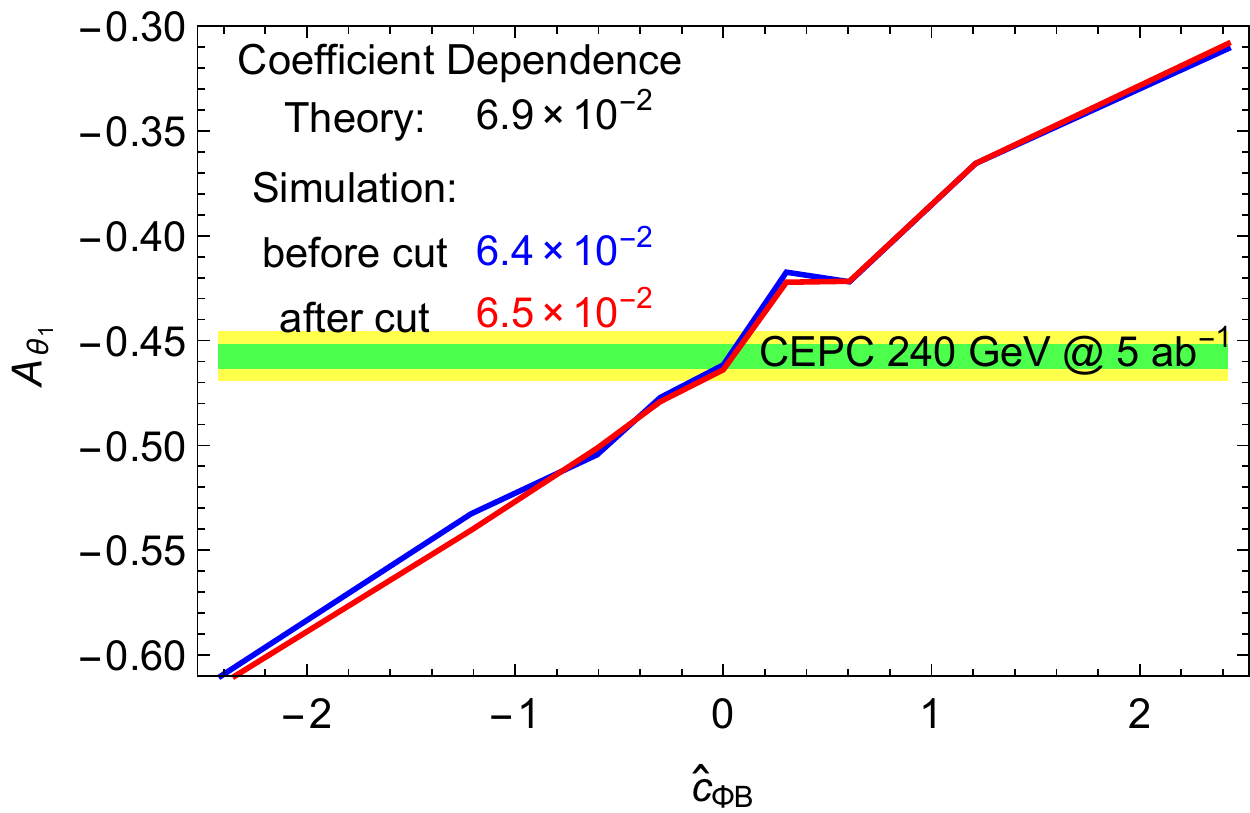}}
\subfigure{
\includegraphics[width=0.475\textwidth]{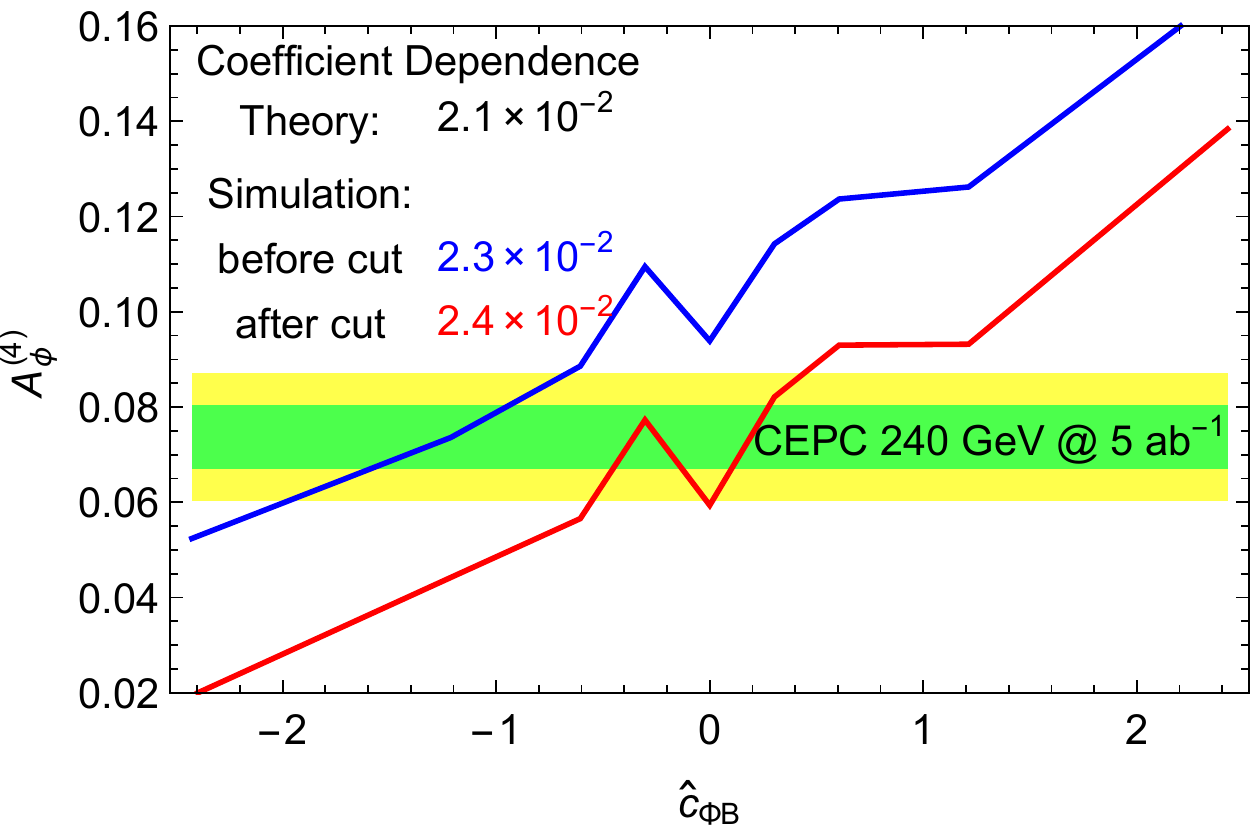}}
\caption[]{Asymmetry observable $A_{\theta1}$ (left panel) and $A_\phi^{(4)}$ (right panel) as a function of the coefficient of operator $O_{\Phi B}$. The blue and red lines indicate the simulated results at parton level before and after cuts. 
The green and yellow band is projected precision for corresponding observable with CEPC 5 $\abi$, assuming measured values follow SM. The numbers are the slope of the asymmetry observable with respect to coefficients $\hat c_{\Phi B}$ at parton level, after cuts and in theory, respectively.}
\label{fig:cuts}
\end{figure}

The signal selection cuts also shift the central value of the observable asymmetries, which likewise effects their diagnostic power in the event of a BSM contribution. In Fig.~\ref{fig:cuts} we show the asymmetry observable and cross section values resulting from our analytical calculation; our simulation before cuts; and our simulation after cuts, in each case following the input parameter choices discussed in Section \ref{sec:observables}. 
To demonstrate the impact of realistic cuts on different values of the dimension-6 operator coefficients, we choose to plot these predictions as a function of $\hat c_{\Phi B}$, the coefficient one of the higher dimensional operators only constrained by precision Higgs measurements. The operator $\CO_{\Phi B}$ affects all CP-even observables $\sigma$, $A_{\theta_1}$, $A_\phi^{(3)}$, $A_\phi^{(4)}$ and $A_{c\theta_1,c\theta_2}$, and is a reasonable proxy for the impact of cuts on various BSM contributions to the angular asymmetries. 

We can see in Fig.~\ref{fig:cuts} that our simulation before cuts agrees well with our analytical calculation, and the cuts alter the asymmetries consistently for different values of $\hat c_{\Phi B}$.  The central values of the asymmetry variables for different values of the coefficients $\hat c_{\Phi B}$ before and after cuts are shown in blue and red lines.  Unsurprisingly, the asymmetries change as a result of the cuts, ranging from sub-percent level for most asymmetry variables to around $40\%$ (reduction) for $A_{\phi}^{(4)}$. 
 Crucially, the slopes of the asymmetries as a function of the Wilson coefficient do not change much before and after cuts. This allows us greatly simplify our forecasting, as it demonstrates that realistic cuts do not alter how asymmetry observables respond to new operators in the regime of interest. 

To study the impact of realistic cuts and detector acceptance on asymmetry observables
we compare our {\tt Madgraph5} analysis result with the data sample from the CEPC Pre-CDR study ground for the SM expectations. Their sample data is generated using {\tt Whizard} \cite{Kilian:2007gr,Moretti:2001zz} for the process of $e^+ e^- \rightarrow H Z$, with $H \rightarrow b \bar{b}$ at $\sqrt{s} = $ 240 GeV. A fast simulation of detector effects using the energy resolution and acceptance in polar angle of the CEPC detector is applied as well.
Our study with the {\tt Madgraph5} sample and with the CEPC preCDR sample agree well in the values of the asymmetries under selection cuts, after taking into account the difference caused by different SM input parameters.
With the initial state radiation the initial state $\ee$ and the intermediate Higgs boson and $Z$ boson are no longer coplanar, causing ambiguities in the definition of asymmetries. We compare the asymmetry values obtained in the lab frame and the $\ee\to ZH$ collision center of mass frame. We find the difference are negligibly small for the precisions of symmetries at CEPC. 

Following our simulation study and the validation of CEPC pre-CDR data sample, we judiciously apply a universal 5\% penalty factor for the sensitivities of asymmetry observables to Wilson coefficients under consideration. This penalty factor is to account for detector effects and other systematics.

In Fig.~\ref{fig:cuts} we also show the one- and two-sigma bands (green and yellow) indicating the constraint that can be placed at CEPC assuming a Standard Model-like central value.  Here we see that the angular variable $A_{\theta_1}$ places an order-of-magnitude stronger bound on $O_{\Phi B}$ compared to $A_{\phi}^{(4)}$. This is unsurprising given our discussion in Section \ref{sec:observables}, but also owes in part to the different precision attainable in the two asymmetries after cuts. We reserve a more complete discussion of the constraints on Wilson coefficients for Section \ref{sec:applications}.

Apart from the systematic impact of cuts and acceptance discussed above, there are several other sources of uncertainty to consider in a realistic treatment of angular asymmetries at future colliders:

\begin{itemize}
\item There are instrumental uncertainties related to uncertainty in the integrated luminosity which affect the cross section measurement, but this cancels for asymmetry observables by construction. 

\item Instrumental uncertainty from beam energy resolution affects the cross section measurement, but this can be calibrated well by other processes, and its effect on asymmetry observables is very small due to the weak dependence on center-of-mass energy exhibited  in \cite{Beneke:2014sba}.

\item Instrumental uncertainty from initial-state-radiation\footnote{We ignore the beamstrahlung effect here, following the prescription of CEPC preCDR study~\cite{CEPCPreCDR}.} affects the reconstruction of the scattering plane, but we have verified that the size of this effect is small at the level of our {\tt Whizard} simulation and is well within the size of our conservatively-assigned overall uncertainty. 
To verify this in detail, we use the signal events for the same process from the CEPC Pre-CDR study ground with initial-state-radiation, reconstruct the angular observables ignoring such effects (i.e., assuming lab frame and c.m. frame are identical) and compare the values of these observable with the true asymmetry using the radiated photon information. Uncertainties resulting from these effects are numerically sub-dominant.

\item Instrumental uncertainty from particle reconstruction energy resolution is also small because the leptons can be measured very precisely and our asymmetry observables rely only on the lepton momentum. This may become a more significant uncertainty if asymmetries are constructed from non-leptonic decay products, or if more information from the decay of the Higgs is employed. 

\item Theoretical uncertainties such as uncertainties on input parameters and electroweak corrections are very important for the precision measurement of angular observables. We consider them to be factorizable from other systematics and note that these uncertainties can be significantly improved in the near future. Current estimates place uncertainties due to NLO electroweak corrections at the (sub)percent level. While they may be somewhat higher in angular observables, substantial improvements may be realized by the advent of future Higgs factories. We note that some angular observables such as $A_\phi^{(3)}$ and $A_{c\theta_1,c\theta_2}$ are highly sensitive to the values of input parameters, and the central values of Standard Model predictions can be altered by as large as a factor of three by different choices of SM input parameters. This makes clear the need for a careful future treatment of Standard Model predictions for these observables.
\end{itemize}


\section{Applications} \label{sec:applications}

Given our estimates for the sensitivity attainable in angular observables at future $e^+ e^-$ colliders, we now consider the implications of this sensitivity for a variety of BSM physics scenarios. Broadly speaking, angular observables both improve the overall reach for BSM physics and constrain linear combinations of Wilson coefficients in the dimension-6 HEFT orthogonal to those constrained by coupling measurements alone.

For the purposes of forecasting, we assume the experimental results are SM-like and obtain the expected constraints on new physics using a simple $\chi^2$ fit.  For the sake of concreteness, we focus on the channel $\ee\to Z H\to \ell^+ \ell^-\,b\bar{b}$ at CEPC with $\sqrt{s} = 240$~GeV and $5\, {\rm ab}^{-1}$ integrated luminosity, although we also forecast sensitivity for several scenarios at FCC-ee.  As we have justified in Section~\ref{sec:actual}, statistical uncertainties dominate for the angular observables in this channel.  As such, in this section we neglect systematic uncertainties and consider only statistical uncertainties based on a sample size of 22100 events (the expected number of events collected by CEPC after selection cuts with $5\, {\rm ab}^{-1}$ integrated luminosity). To compensate the omission of systematics, we judiciously apply a universal $5\%$ penalty factor for the sensitivities of asymmetry observables to Wilson coefficients, as mentioned in Section~\ref{sec:actual}.  For the uncertainty in the cross section, we adopt the values in the preCDR \cite{CEPCPreCDR}, which are $0.9\%$ for the $\mm b\bar{b}$ channel and $1.1\%$ for the $\ee b\bar{b}$ channel.  The combined precision for the $\ee\to Z H\to \ell^+ \ell^-\,b\bar{b}$ channel is therefore $0.7\%$, assuming statistical uncertainties dominate.

In what follows we will also neglect theory uncertainties in the Standard Model predictions for the total cross section and for angular observables. Recent forecasts for precision measurements of the associated production cross section suggest that uncertainties on the order of O(0.5\%) are realistic. For angular observables the situation is somewhat less clear, but we expect substantial progress in the study of NLO corrections to angular variables in anticipation of a future $e^+ e^-$ collider.

The $\chi^2$ from the rate measurements ($\chi^2_{\rm rate}$), the angular measurements ($\chi^2_{\rm angles}$) and the cominbation of all measurements ($\chi^2_{\rm total} $) are defined as\footnote{It is unfortunate that the conventions for the cross section and the standard deviation are both $\sigma$.  Here to distinguish the two we denote the total cross section as $X$.} 
\begin{align}
\chi^2_{\rm rate} = &~ \frac{(X_{\rm NP}-X_{\rm SM})^2}{\sigma^2_X} \,,  \label{eq:chix}\\
\chi^2_{\rm angles} = &~ \sum_i \frac{(\mathcal{A}^i_{\rm NP}-\mathcal{A}^i_{\rm SM})^2}{\sigma^2_{\mathcal{A}^i}} \,, \label{eq:chia}\\
\chi^2_{\rm total} = &~ \chi^2_{\rm rate} + \chi^2_{\rm angles} \,, \label{eq:chixa}
\end{align}
where $X_{\rm SM}$ and $\mathcal{A}^i_{\rm SM}$ are the ``measured" values which we assume to be SM-like; $X_{\rm NP}$ and $\mathcal{A}^i_{\rm NP}$ are the predictions of new physics, which can be written as functions of the Wilson coefficients as {\it e.g.} in Eq.~(\ref{eq:csn})--(\ref{eq:Ac12n}); and the $\mathcal{A}^i$ are summed over $\mathcal{A}_{\rm  \theta_1}$, $\mathcal{A}_{\rm  \phi}^{(1)}$, $\mathcal{A}_{\rm  \phi}^{(2)}$, $\mathcal{A}_{\rm \phi}^{(3)}$, $\mathcal{A}_{\rm \phi}^{(4)}$ and $\mathcal{A}_{c \theta_1, c \theta_2}$. Here $\sigma_X$ and $\sigma_{\mathcal{A}^i}$ are the $1\sigma$ uncertainties for the rate and angular observables, respectively.  We have also neglected the correlations among (the experimental measurements of) the observables, which we expect to be small.

The formulation in Eq.~(\ref{eq:chix}--\ref{eq:chixa}) is also applicable to the discrimination of different new physics cases, for which one could assume the measured values of the observables are given by some benchmark new physics scenario.  The changes in the expected precisions are negligible, unless the NP predictions are dramatically different from the SM ones (in which case the effective theory description will break down).  Since there is no additional information, we will focus on forecasting in the scenario that the experimental results are SM-like.

The rest of this section is organized as follows:  In Section~\ref{sec:cwc}, we present the constraints on the Wilson coefficients in the Higgs effective Lagrangian with a model-independent approach.  In Section~\ref{sec:hzr}, we discuss in detail the constraints on the $HZ\gamma$ coupling, for which angular observables provide sensitivity comparable to that of direct measurement at $e^+ e^-$ colliders.  In Section~\ref{sec:stop}, we demonstrate how the constraints from angular observables can be applied to specific models of new physics, using light stops as an example.

\subsection{Constraining Wilson coefficients}
\label{sec:cwc}

\begin{figure}[htp!]
\centering
\includegraphics[width=0.31\textwidth]{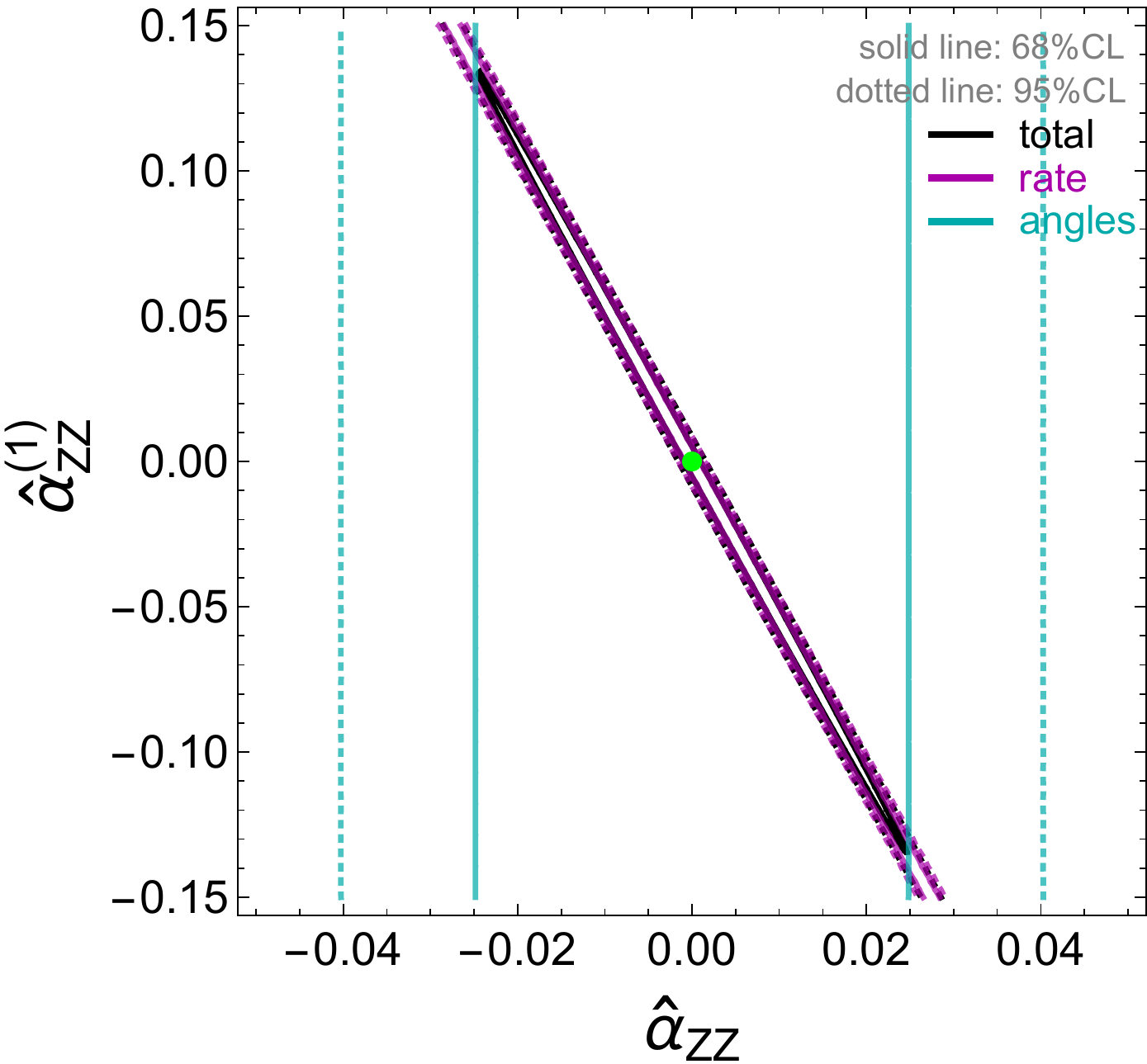}  \hspace{0.1cm}
\includegraphics[width=0.31\textwidth]{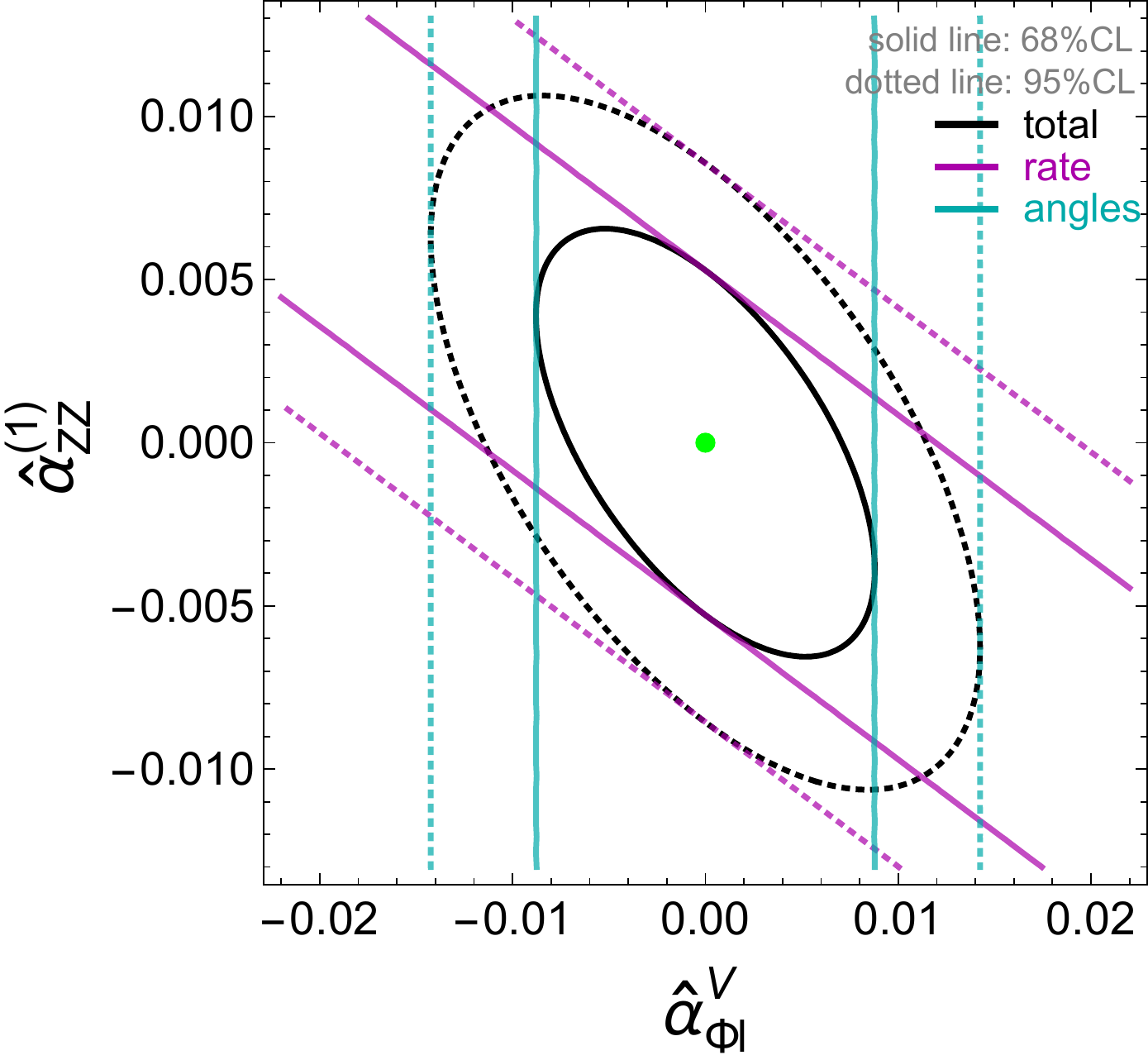}  \hspace{0.1cm}
\includegraphics[width=0.30\textwidth]{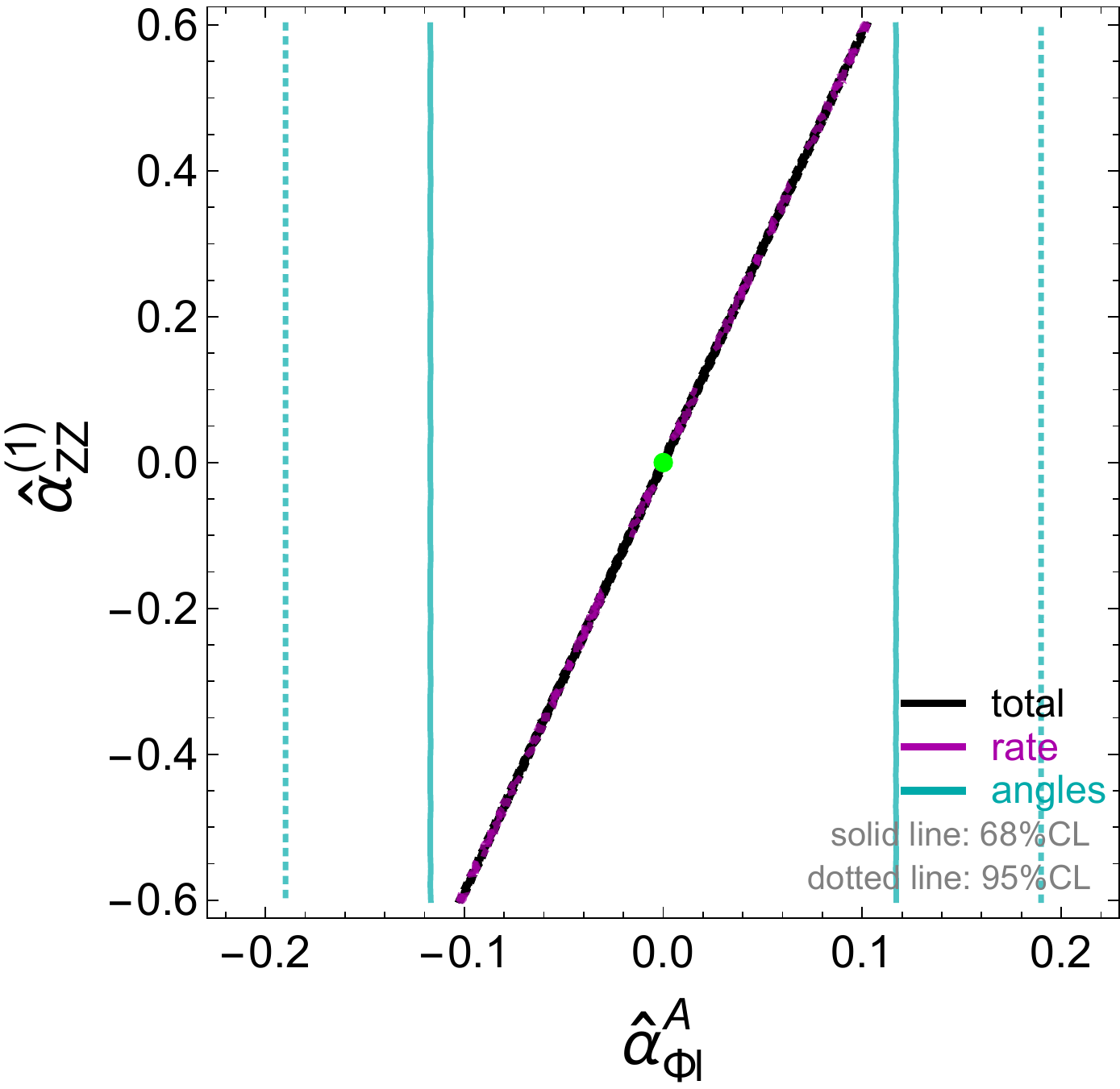}  \\ 
\vspace{0.2cm}
\includegraphics[width=0.31\textwidth]{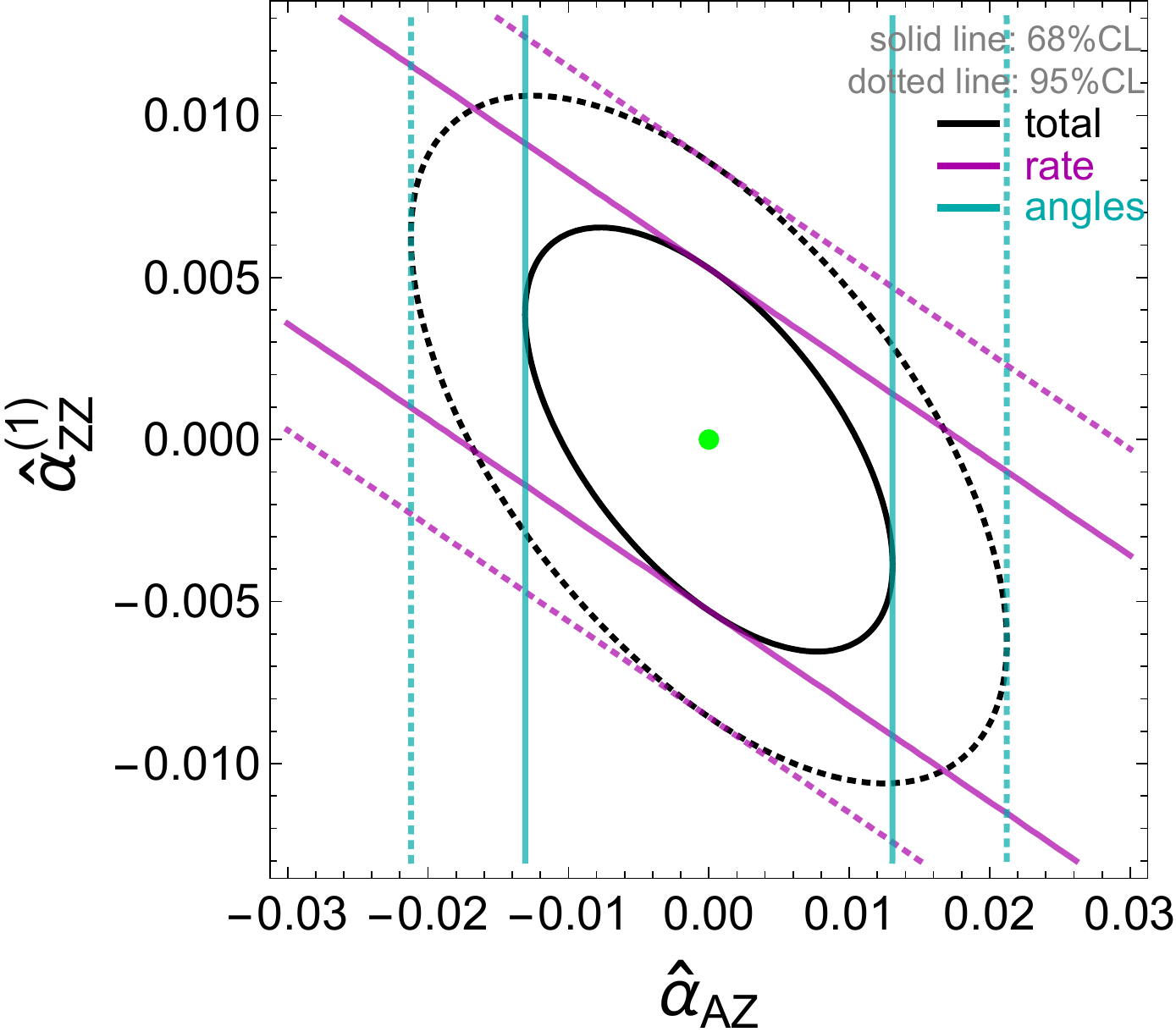}  \hspace{0.1cm}
\includegraphics[width=0.31\textwidth]{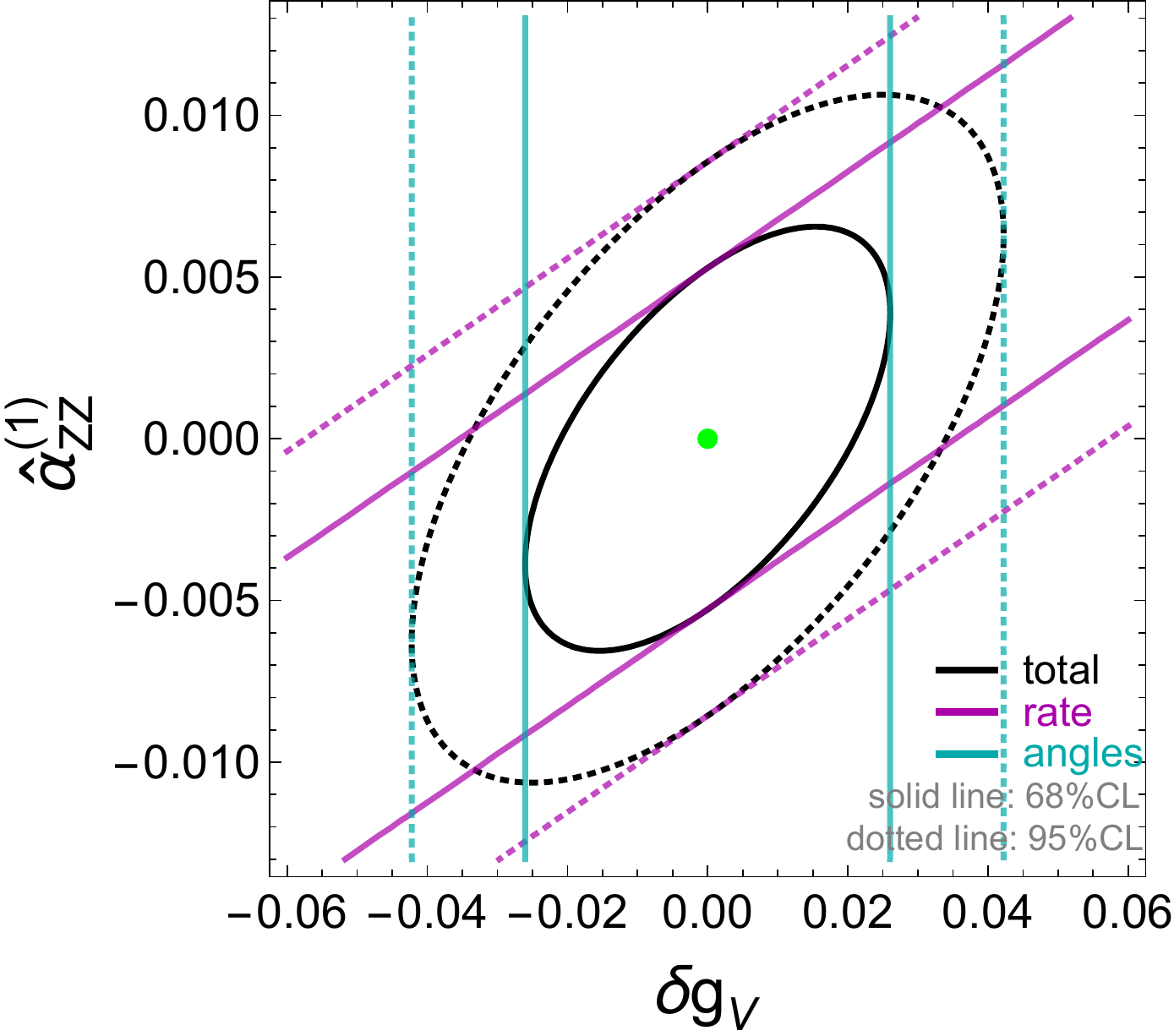}  \hspace{0.1cm}
\includegraphics[width=0.30\textwidth]{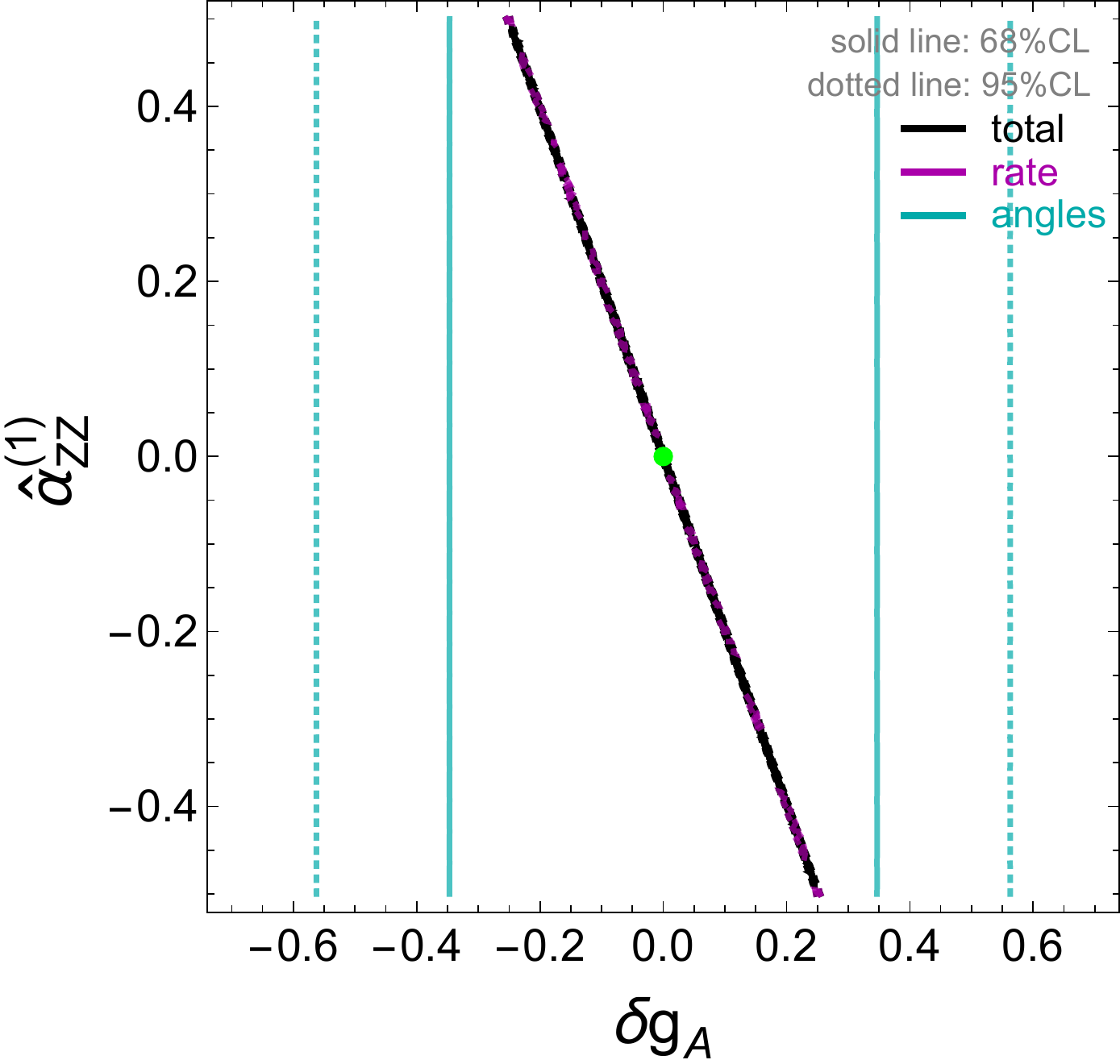}  \\ 
\vspace{0.2cm}
\includegraphics[width=0.31\textwidth]{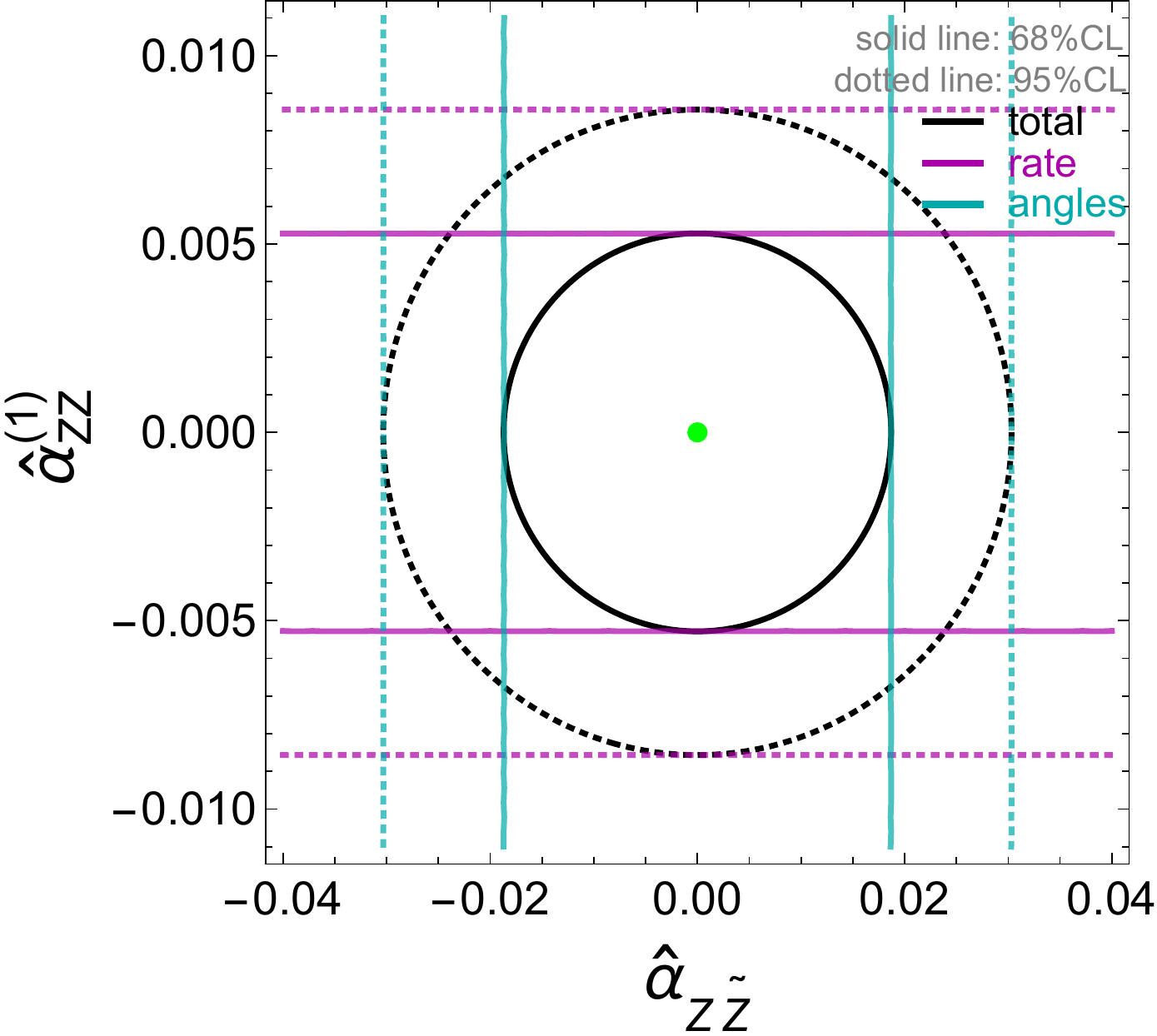}  \hspace{0.1cm}
\includegraphics[width=0.31\textwidth]{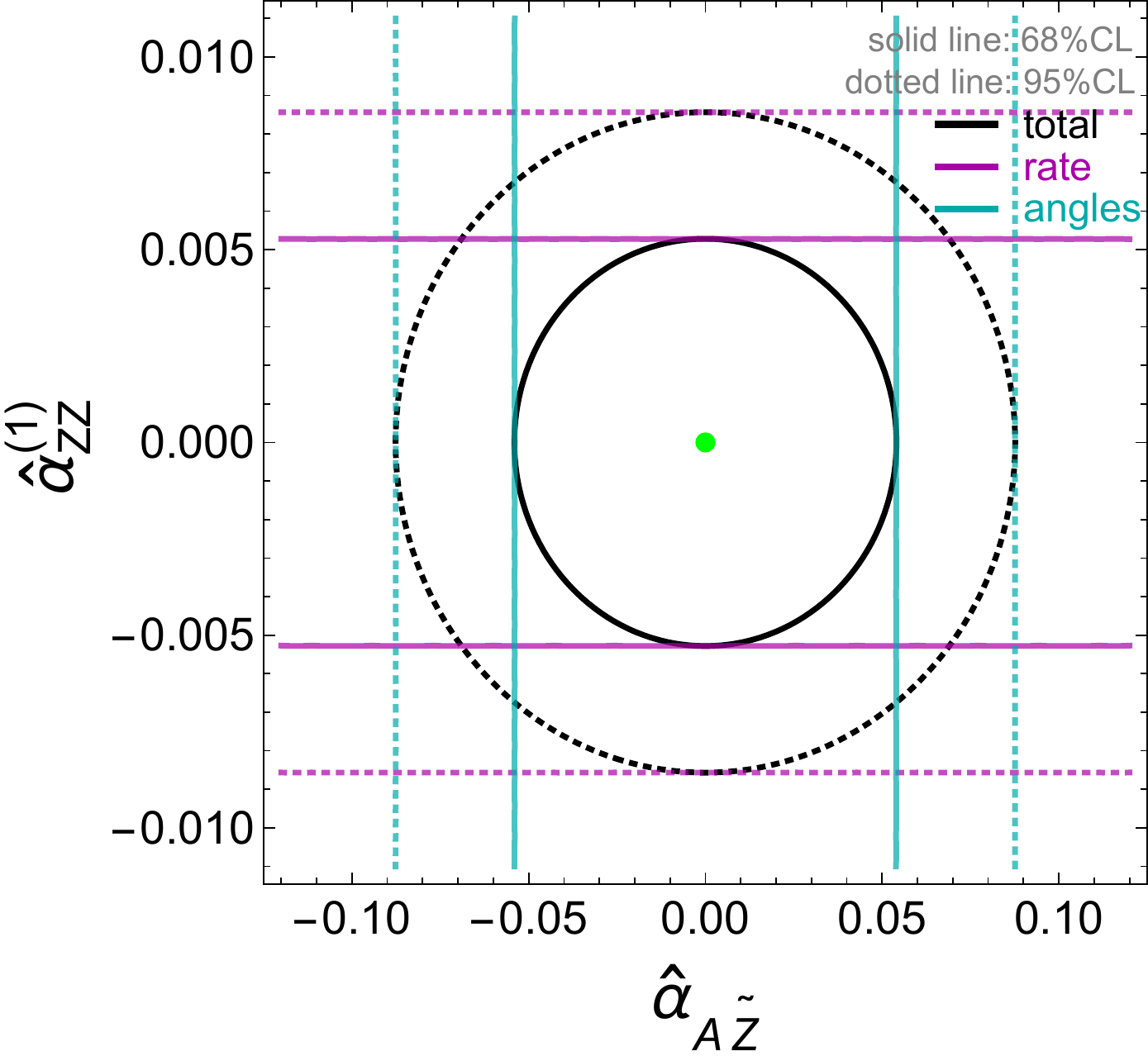}  
\caption{Expected constraints from the $\ee\to Z H\to \ell^+ \ell^-\,b\bar{b}$ process at the CEPC with $\sqrt{s} = 240$~GeV and $5\, {\rm ab}^{-1}$ data, assuming SM-like results.  Each plot shows the constraints for $ \ha^{(1)}_{ZZ}$ and one of the other Wilson coefficients in Eq.~(\ref{eq:wclist}), while the rest are set to zero.  The purple contours show the constraints from the rate measurements only, the cyan contours show the constraints from measurements of angular observables (combined) only, and the black contours show the total combined constraints from both rate and angular measurements.  The solid (dotted) lines corresponds to $68\%$($95\%$) confidence level (CL).  The green dot at $(0,0)$ indicates the SM prediction. }
\label{fig:cwc}
\end{figure}

In this section we present the model-independent constraints on the Wilson coefficients in the Higgs effective Lagrangian, Eq.~(\ref{eq:efflag}), parameterized the 9 Wilson coefficients in Eq.~(\ref{eq:wclist}), which are  
\begin{equation}
\ha_{ZZ} \,, ~~~ \ha^{(1)}_{ZZ} \,, ~~~  \ha^V_{\Phi \ell} \,, ~~~ \ha^A_{\Phi \ell} \,, ~~~ \ha_{AZ} \,, ~~~ \delta g_V \,, ~~~ \delta g_A \,, ~~~  \ha_{Z\widetilde{Z}} \,, ~~~  \ha_{A\widetilde{Z}} \,.  \nonumber
\end{equation}
Treating the 9 coefficients as independent parameters, there are totally 7 constraints from the rate and the six asymmetry observables, less than the number of unknowns.  Therefore, one cannot obtain independent constraints on the Wilson coefficients without making further assumptions.  However, with a reduced set of coefficients the angular observables can break the degeneracy of the rate measurement, which by itself could only constrain one linear combination of the Wilson coefficients.  To illustrate this point, we focus on two coefficients at a time while setting the rest to zero.  One of the coefficients is always chosen to be $\ha^{(1)}_{ZZ}$, which parameterizes a modification of the SM $HZ^\mu Z_\mu$ interaction and is most strongly constrained by the rate measurement.  The angular observables, being normalized to the total cross section, are independent of $\ha^{(1)}_{ZZ}$ by construction.  In Fig.~\ref{fig:cwc}, we show the expected constraints in the two-dimensional parameter space consisting of $\ha^{(1)}_{ZZ}$ and one of the remaining coefficients, assuming SM-like measurements.   
The total combined constraints from both rate and angular measurements are shown along with the ones from the rate measurements or the (combined) angular observables alone.  From Fig.~\ref{fig:cwc}, it is clear that the rate measurement alone only constrains a linear combination of the two coefficients. The inclusion of angular observables typically allows considerable discrimination between Wilson coefficients. 
In particular, in Fig.~\ref{fig:cwc} it is apparent that angular observables have appreciable discriminating power for the coefficients $\ha^V_{\Phi \ell}, \ha_{AZ}, \delta g_V,  \ha_{Z\widetilde{Z}},$ and $\ha_{A\widetilde{Z}}$. Note that the angular obervables $\mathcal{A}^{(1)}_\phi, \mathcal{A}^{(2)}_\phi$ depend only on CP-odd operators, which are zero in the Standard Model and thus entirely dominated by contributions from the dimension-6 EFT. The remaining angular observables depend only on CP-even operators, which generally accumulate radiative contributions in the Standard Model.

In principle, precision $e^+ e^-$ colliders could be used to set appreciable bounds on CP-odd operators via measurements of the observables $\mathcal{A}^{(1)}_\phi, \mathcal{A}^{(2)}_\phi$. However, these operators are much more strongly constrained by bounds on the electron EDM, as they contribute at one-loop order through Barr-Zee type diagrams \cite{Barr:1990vd}. In \cite{Fan:2013qn} the corresponding bounds on $\CO_{\Phi B},$ $\CO_{\Phi W}$, and $\CO_{\Phi WB}$ were computed using $|d_e| / e < 1.05 \times 10^{-27}$. Recently the ACME experiment improved this bound to $|d_e| / e < 8.7 \times 10^{-29}$ \cite{Baron:2013eja}. Accounting for the improved limit, for $\Lambda = 1$ TeV, this corresponds to bounds of $\alpha_{\Phi B} < 1.4 \times 10^{-4}$, $\alpha_{\Phi W} < 3.6 \times 10^{-5}$, and $\alpha_{\Phi WB} < 6.7 \times 10^{-5}$ when the operators are considered independently. For $\alpha_i = 1$, this equivalently corresponds to bounds on the scales of the dimension-6 operators $\CO_{\Phi B},$ $\CO_{\Phi W}$, and $\CO_{\Phi WB}$ of $\Lambda > 83, 167,$ and 122 TeV. Thus we conclude that potential limits on $\mathcal{A}^{(1)}_\phi, \mathcal{A}^{(2)}_\phi$ at $e^+ e^-$ colliders, while appreciable, are far from competitive with existing bounds coming from the electron EDM.

While sensitivity to the CP-odd observables is not competitive with EDM experiments, sensitivity to $\ha^V_{\Phi \ell}, \ha_{AZ}$, and $\delta g_V$ provides meaningful improvement over bounds from rate measurements alone. We will discuss the particular utility of the $\ha_{AZ}$ constraint in the next subsection.

Of course, the combination of angular observables is not universally useful, and in some cases (namely $\ha_{ZZ}$, $\ha^A_{\Phi \ell}$ and $\delta g_A$), 
the inclusion of angular observables provides little additional information relative to the rate measurement.  
It should be noted that for these cases we have chosen quite large plot ranges for the sake of illustration; in reality, for large Wilson coefficients the effective theory description would break down.

In addition to the constraints in two-dimensional parameter spaces, we provide in Table~\ref{tab:cwc1} the constraints on individual Wilson coefficients with the assumption that all other coefficients are zero.    Table~\ref{tab:cwc1} shows the $1\sigma$ uncertainties for each Wilson coefficient (setting others to zero) from the rate measurements only, the angular observables measurements only, and the combination of the two.  We use ``$\infty$" to denote coefficients for which no constraint can be derived within our procedure.  In particular, the angular observables are insensitive to $\ha^{(1)}_{ZZ}$ by construction, while the rate measurements are independent of the CP-odd operators at leading order in the Wilson coefficients.

\begin{table}[h]
\centering
\begin{tabular}{|c||c|c|c|c|c|c|c|c|c|c|} \hline
   & $\ha_{ZZ}$    &  $\ha^{(1)}_{ZZ}$  & $ \ha^V_{\Phi \ell}$ & $\ha^A_{\Phi \ell}$ & $\ha_{AZ}$ & $\delta g_V$  & $\delta g_A$   & $\ha_{Z\widetilde{Z}}$ &  $\ha_{A\widetilde{Z}}$    \\  \hline\hline
 rate                &  0.00064    &  0.0035    &  0.0079   &  0.00059  &  0.012   &  0.023  &  0.0018  & $\infty$  &  $\infty$     \\ \hline
 angles            &  0.016    &   $\infty$     &  0.0058     &  0.078   &  0.0087   &  0.017     &  0.23     & 0.012    &  0.036    \\ \hline\hline
 total                &  0.00064    &   0.0035    &   0.0047    &  0.00059  &  0.0070   &   0.014    &   0.0018    &  0.012  &  0.036    \\ \hline
\end{tabular}
\caption{$1\sigma$ uncertainties for individual Wilson coefficients, with the assumption that all other coefficients are zero.  The second row shows the constraints from the rate measurements only, the third row shows the constraints from measurements of angular observables (combined) only, and the last row shows the total combined constraints from both rate and angular measurements.  If no constraint could be derived within our procedure, a $\infty$ is shown.}
\label{tab:cwc1}
\end{table}

\begin{figure}[htp!]
\centering
 \includegraphics[width=0.31\textwidth]{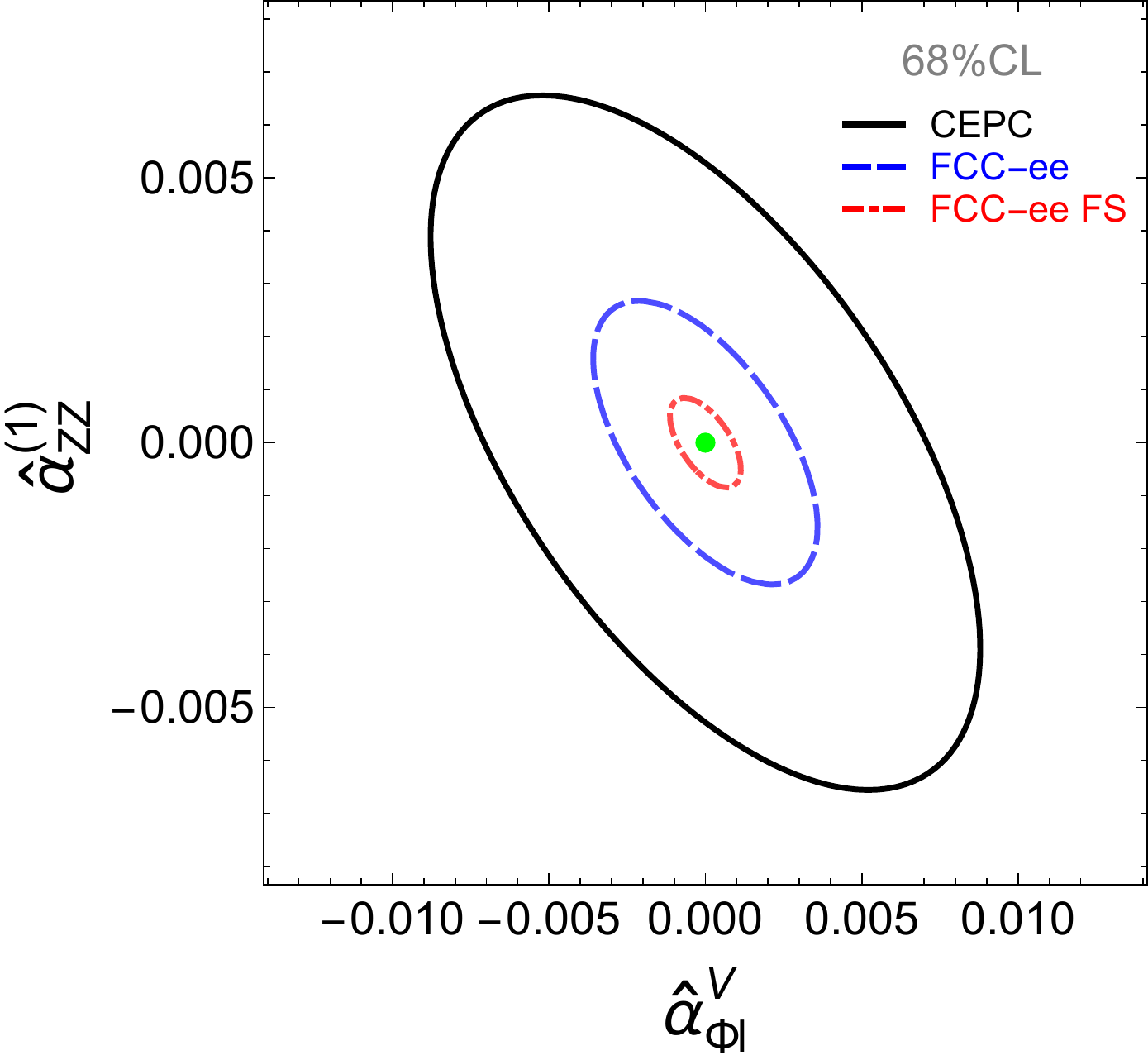}  \hspace{0.1cm} 
 \includegraphics[width=0.325\textwidth]{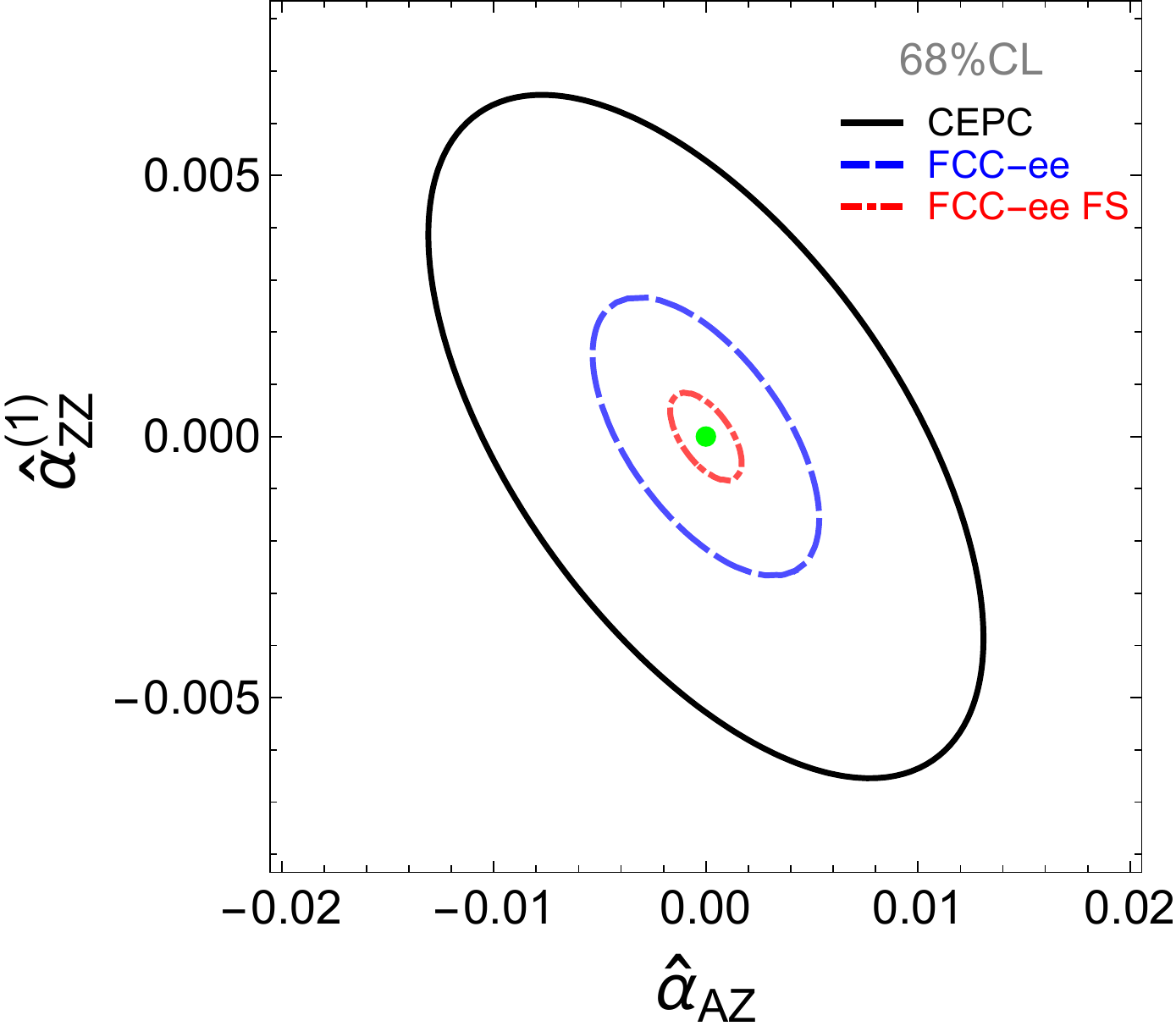}  \hspace{0.1cm}
 \includegraphics[width=0.315\textwidth]{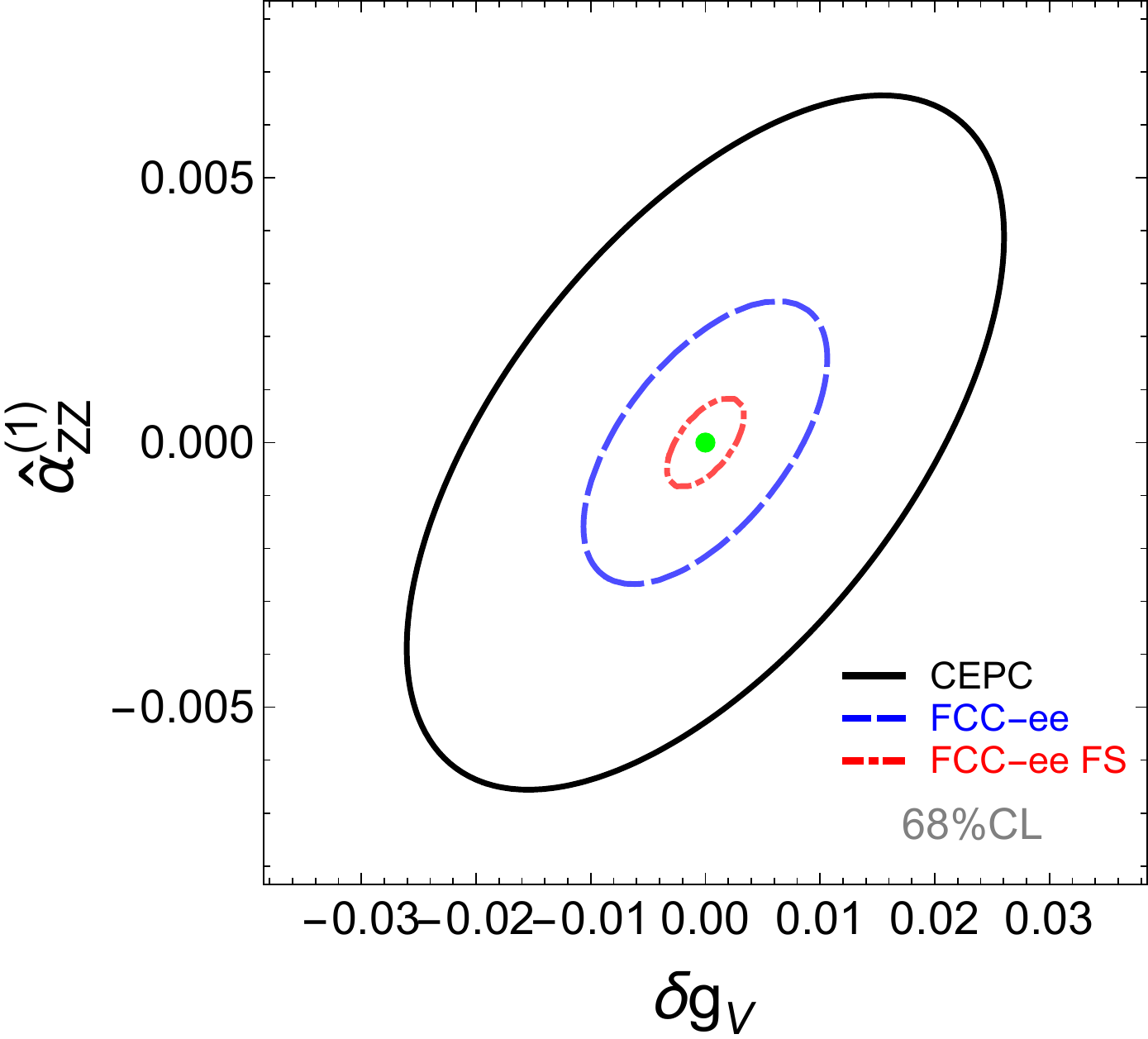} 
\caption{Expected constraints on the Wilson coefficients for different collider scenarios, assuming SM-like results.  The constraints from the measurements of rate  and angular observable are combined together.  Each plot shows the constraints for $ \ha^{(1)}_{ZZ}$ and one of the other Wilson coefficients in Eq.~(\ref{eq:wclist}), while the rest are set to zero.  The black contour shows the constraints for {\bf CEPC}, the blue dashed contour shows the constraints for {\bf FCC-ee}, and the red dot-dashed contour shows the constraints for {\bf FCC-ee FS}.  The three scenarios all use the $\ee\to Z H\to \ell^+ \ell^-\,b\bar{b}$ process at $\sqrt{s} = 240$~GeV, with a sample size of 22100, 132600 and 1326000 events for CEPC, FCC-ee and FCC-ee FS, respectively.  All contours correspond to $68\%$ CL.  The green dot at $(0,0)$ indicates the SM prediction.}
\label{fig:awc}
\end{figure}

As discussed in Section~\ref{sec:actual}, with the same running time FCC-ee is able to deliver a sample size 6 times larger than that of CEPC.  It is also reasonable to expect that statistical uncertainties dominate for the  $\ee\to Z H\to \ell^+ \ell^-\,b\bar{b}$ process at FCC-ee as they do at CEPC.  Furthermore, the inclusion of additional decay modes of $H$ and $Z$ would increase the statistics and could potentially significantly increase the constraining power.  While the reaches of other channels would require further study, to illustrate their potential usefulness we perform a naive scaling of statistics from the FCC-ee $\ee\to Z H\to \ell^+ \ell^-\,b\bar{b}$ process by another factor of 10, and denote this scenario as {\bf FCC-ee FS} (full statistics).  A comparison of the reaches of the three scenarios, {\bf CEPC}, {\bf FCC-ee} and {\bf FCC-ee FS}, is shown in Fig.~\ref{fig:awc} for a selection of Wilson coefficients. Here we use the  $\ee\to Z H\to \ell^+ \ell^-\,b\bar{b}$ process at $\sqrt{s} = 240$~GeV with sample sizes of 22100, 132600 and 1326000 events for CEPC, FCC-ee and FCC-ee FS, respectively.    

\subsection{The $HZ\gamma$ coupling}
\label{sec:hzr}

As we have seen, angular observables in $e^+ e^- \to ZH$ provide an additional means of probing $\ha_{AZ}$, thereby constraining  anomalous contributions to the $hZ\gamma$ coupling. To date, much attention has been devoted to constraining the $hZ\gamma$ coupling in decays $h \to Z \gamma$ (which can be observed at the LHC) and in the production mode $e^+ e^- \to \gamma h$ (which can be observed at future $e^+ e^-$ colliders). However, the contribution of the $hZ\gamma$ to $e^+ e^- \to ZH$ via an intermediate photon provides a complementary probe at $e^+ e^-$ colliders that is not sensitive to the potentially complicated backgrounds faced by $e^+ e^- \to \gamma h$.

This is perhaps not surprising. In general, one expects angular observables to provide a powerful handle on small deviations in the $h Z \gamma$ coupling, insofar as BSM contributions appear at $\mathcal{O}(\alpha_{AZ})$ via interference with the tree-level SM process $e^+ e^- \to ZH$, whereas in $e^- e^- \to \gamma h$ they arise either directly at $\mathcal{O}(\alpha_{AZ}^2)$ or at $\mathcal{O}(\alpha_{AZ})$ via interference with the {\it loop-level} SM process $e^+ e^- \to \gamma h$.\footnote{We note that our analysis is restricted to tree-level processes in the Standard Model, while the SM contribution to the $hZ\gamma$ coupling arises at one loop. Including the SM contribution would not substantially alter our analysis, as it only shifts the central value of the angular observables. }

The limits on $h Z \gamma$ that may be obtained by precision measurement of both the rate and angular distributions of $e^+ e^- \to ZH$ ($\to \ell^+ \ell^-\,b\bar{b}$) is shown in 
Table~\ref{tab:cwc1}, amounting to $-0.0070 \leq \widehat{\alpha}_{AZ} \leq 0.0070$  at CEPC.  By comparison, direct measurement of the $e^+ e^- \to \gamma h$ process at CEPC (scaling up the result of \cite{Cao:2015iua} to 5 ab$^{-1}$) leads to a projected bound of $-0.008 \leq \widehat{\alpha}_{AZ} \leq 0.003$. While this is somewhat better than the projected bound from $e^+ e^- \to ZH$, the analysis in \cite{Cao:2015iua} includes only the backgrounds from hard processes $e^+ e^- \to \gamma b \bar b$. In general one expects additional backgrounds from beamstrahlung that are more difficult to characterize and may further complicate the direct measurement. At the very least, it is clear that angular observables provide a surprising and competitive avenue for probing the $h Z \gamma$ coupling at future $e^+ e^-$ colliders.

\subsection{Stops}
\label{sec:stop}

As a final example of the discriminating power of angular observables, we consider a concrete weakly-coupled model that may be constrained with precision measurements at $e^+ e^-$ colliders: scalar top partners (stops) in supersymmetric extensions of the Standard Model. For simplicity, we will consider stops with degenerate stop soft masses $m_{\tilde t}^2 = m_{\tilde Q_3}^2 = m_{\tilde t_R}^2$ plus mixing terms of the form $X_t = A_t - \mu \cot \beta$. The mass scale of the effective operators is $\Lambda = m_{\tilde t}$.  Wilson coefficients for this scenario were computed in \cite{Henning:2014gca}, while the constraint on the stop parameter space due to rate measurements at $e^+ e^-$ colliders was determined in \cite{Craig:2014una} \footnote{See also \cite{Drozd:2015kva, Drozd:2015rsp} for similar studies.}. Here we include the additional sensitivity contributed by angular observables by translating the results of \cite{Henning:2014gca, Craig:2014una} into our preferred basis of Wilson coefficients and applying the results of the previous section.

\begin{figure}[htp!]
\centering
\includegraphics[width=0.31\textwidth]{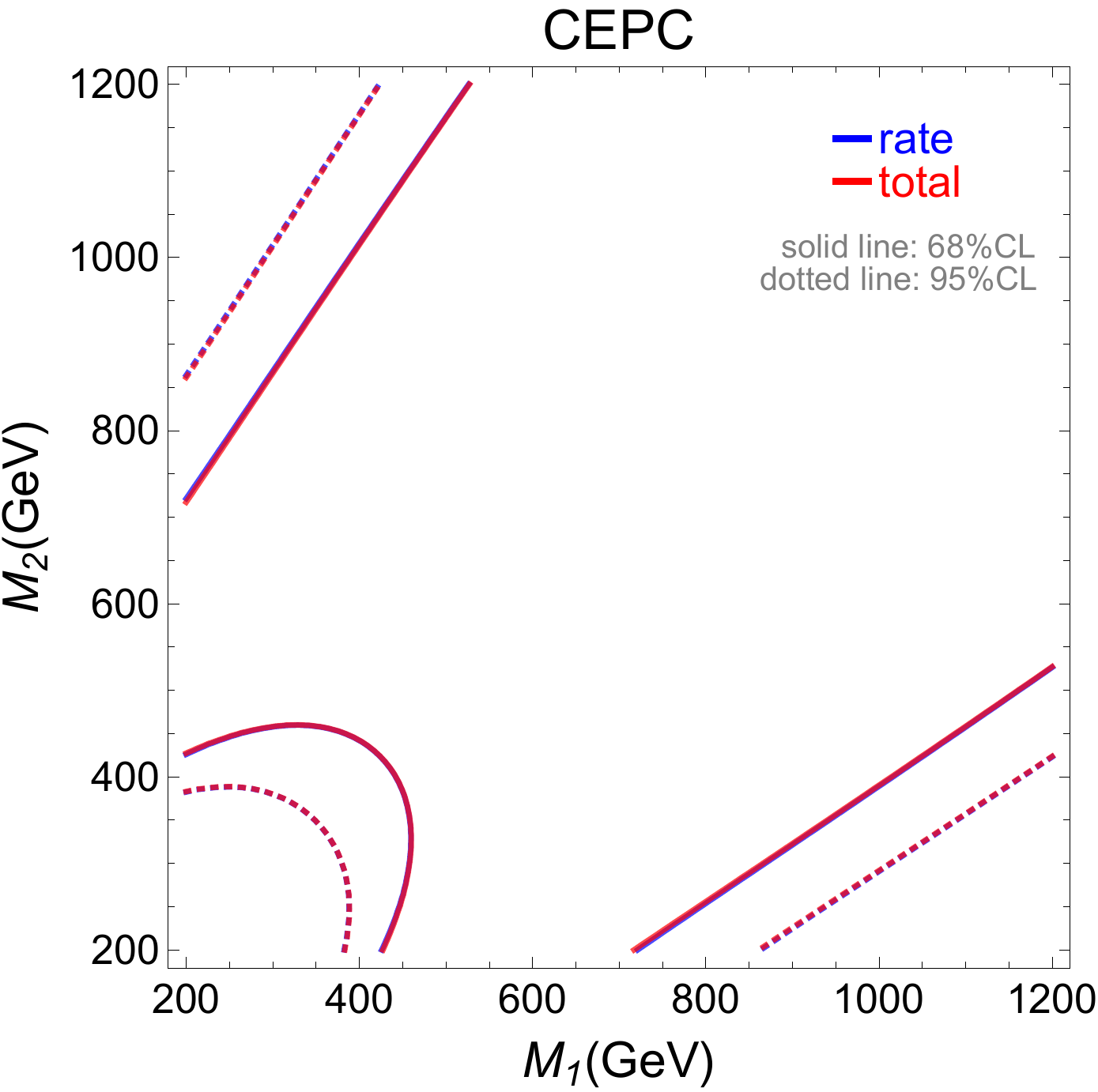}  \hspace{0.3cm}
\includegraphics[width=0.31\textwidth]{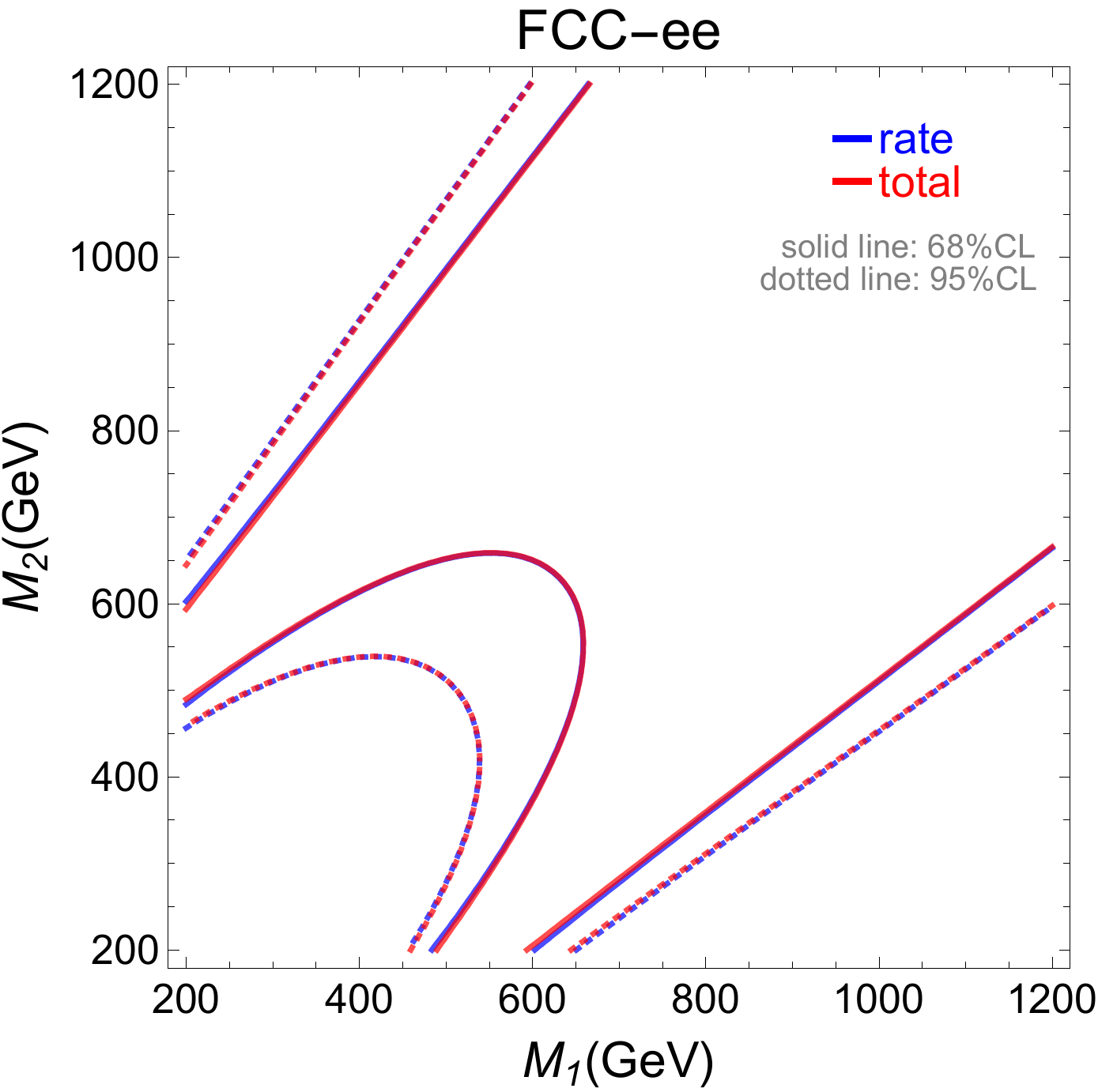}  \hspace{0.3cm}
\includegraphics[width=0.31\textwidth]{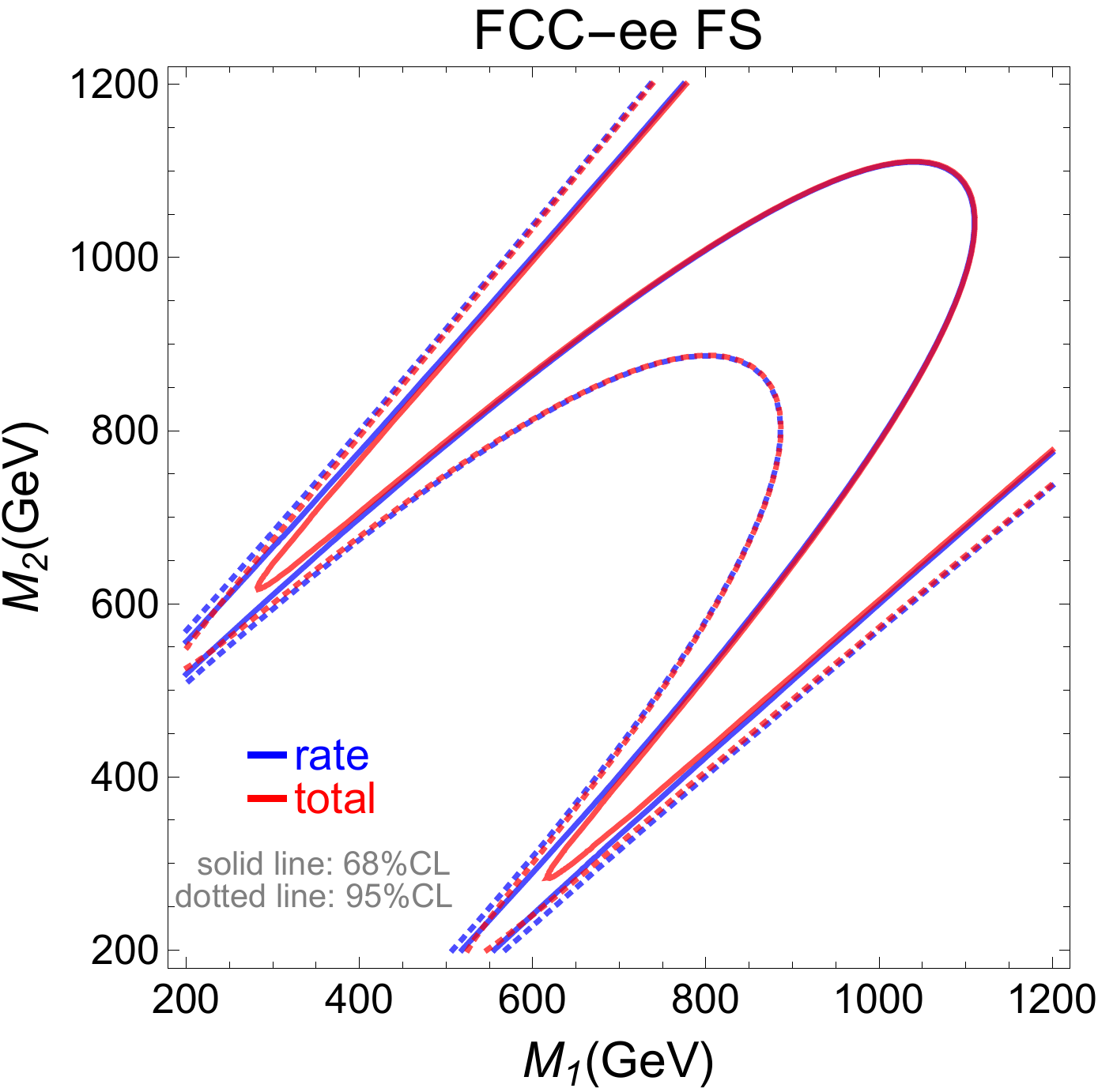}
\caption{ Expected constraints in the $(M_1, M_2)$ plane for different collider scenarios, assuming SM-like results.  $M_1$ and $M_2$ are the two mass eigenvalues of the left- and right-handed stops.  The three scenarios, {\bf CEPC}, {\bf FCC-ee} and {\bf FCC-ee FS} are described in Section~\ref{sec:cwc}. We set $\tan{\beta}=10$.  The blue contours show the constraints from the rate measurements only and the red contours show the total combined constraints from the measurements of rate and the angular observables .  The solid (dotted) lines corresponds to $68\%$($95\%$) CL.  The region in the upper-right part of each plot is allowed by projected coupling measurements.}
\label{fig:stopM2vsM1}
\end{figure}

In Fig.~\ref{fig:stopM2vsM1} we show the sensitivity provided by rate measurements and the inclusion of angular observables in the plane of the two stop mass eigenvalues $M_1$ and $M_2$, which are functions of $m_{\tilde t}^2$ and $X_t$ given by the stop mass mixing matrix.  For definiteness we have set $\tan{\beta}=10$, while the results are insensitive to $\tan{\beta}$ as long as $\tan{\beta} \gtrsim$ few.

The features of the exclusion provided by rate measurements were discussed extensively in  \cite{Craig:2014una}. The most noteworthy feature of the rate measurements is the so-called ``blind spot'' along the line $M_2 = M_1 +\sqrt{2} m_t$ where the shift in the $hZZ$ rate is zero. Such blind spots arise more generally in stop corrections to various Higgs properties such as $hgg$, $h \gamma \gamma$ couplings and precision electroweak observables. Each blind spot corresponds to a zero in physical linear combinations of Wilson coefficients. While the exact zeroes in observables arise in different places in the $M_1 - M_2$ plane, they are collected around the line in which the coupling of the lightest stop mass eigenstate to the Higgs goes to zero. 

In general, the addition of angular observables does not lead to immense improvements over the rate measurement in generic regions of parameter space. This is not surprising, since the relevant Wilson coefficients well-constrained by angular observables are generated at one loop and thus are small for all values of the stop masses. However, it is apparent in  Fig.~\ref{fig:stopM2vsM1} that the addition of angular observables provides significantly improved sensitivity in the blind spot of the $hZZ$ rate measurement. This is simply because the Wilson coefficients contributing to angular observables are suppressed but nonzero along the line where the $ZH$ cross section shift is zero, and so provide complementary sensitivity at small $M_1, M_2$ provided sufficient statistics. This demonstrates the value of angular observables even in the case of BSM scenarios that are generally well-constrained by rate measurements.

\section{Conclusions} \label{sec:conclusions}

Future $e^+ e^-$ provide unprecedented opportunities to explore the Higgs sector. The large sample size of clean Higgs events may be used to constrain not only deviations in Higgs couplings, but also non-standard tensor structures arising from BSM physics. While the former are readily probed by rate measurements, the latter may be effectively probed using appropriately-constructed angular asymmetries. In this work we have initiated the study of angular observables at future $e^+ e^-$ colliders such as CEPC and FCC-ee. We have taken particular care to account for the impact of realistic cut acceptance and detector effects on angular asymmetries. 

Our primary result is a forecast of the precision with which angular asymmetries may be measured at future $e^+ e^-$ colliders. We have translated this forecast into projected sensitivity to a range of operators in the dimension-6 EFT, where angular measurements provide complementary sensitivity to rate measurements. Among other things, we have found that angular asymmetries provide a novel means of probing BSM corrections to the $h Z \gamma$ coupling beyond direct measurement of $e^+ e^- \to h \gamma$. We also apply our results to a complete model of BSM physics, namely scalar top partners in supersymmetric extensions of the Standard Model, where angular observables help to constrain the well-known ``blind spot'' in rate measurements.  

There are a wide range of interesting future directions. In this work we have focused on $ZH$ events with $Z \to \ell^+ \ell^-$ and $h \to b \bar b$ in order to obtain a relatively pure sample of signal events without significant background contamination. Of course, there will be far more events involving alternate decays of the $Z$ and Higgs which, while not background-free, could add considerable discriminating power. It would be useful to conduct a realistic study of these additional channels to determine the maximum possible sensitivity of angular asymmetries. Although we have taken care to account for the impact of cut acceptance and detector effects on angular asymmetries, our work has neglected the possible impact of theory uncertainties in the Standard Model prediction for angular asymmetries. A detailed estimate of current and projected theory uncertainties in the Standard Model prediction for Higgsstrahlung differential distributions would be broadly useful to future studies. More generally, this work serves as a starting point for investigating the full set of Higgs properties accessible at future $e^+ e^-$ colliders.

\section*{Acknowledgements}
We thank Gang Li and Xin Mo for providing us the CEPC pre-CDR simulation events. We thank Nima Arkani-Hamed, Marco Farina, Marat Freytsis, Matthew McCullough, Maxim Perelstein, Matt Reece, and Lian-Tao Wang for useful conversations. KW and JG also thank Cai-Dian L$\rm \ddot{u}$ for assistance. NC is supported by the Department of Energy under the grant DE-SC0014129. NC and ZL  acknowledge the hospitality of the IHEP Center for Future High Energy Physics where this work was initiated. Fermilab is operated by Fermi Research Alliance, LLC under Contract No. DE-AC02- 07CH11359 with the United States Department of Energy. KW and JG are supported in part by the CAS Center for Excellence in Particle Physics (CCEPP).  JG is also supported in part by the Chinese Academy of Science (CAS) International Traveling Award under Grant H95120N1U7.

\appendix

\section{Angular coefficients} \label{app:j}

For completeness, here we list the various $J_i$ coefficients computed first in \cite{Beneke:2014sba}. These coefficients are conveniently expressed in terms of components of the matrix element as
\begin{eqnarray}
 J_1 &=& 2  r s \left(g_A^2+g_V^2\right) \left(|H_{1,V}|^{2}+|H_{1,A}|^{2}\right) \\
J_2 &=& \kappa \left(g_A^2+g_V^2\right) \left[\kappa \left(|H_{1,V}|^{2 }+|H_{1,A}|^{2}\right)+\lambda {\rm Re} \left( H_{1,V} H^{*}_{2,V} + H_{1,A} H^{*}_{2,A}\right) \right] \\
J_3 &=& 32 r s g_A g_V {\rm Re} \left ( H_{1,V}  H_{1,A}^{*} \right ) \\
J_4 &=& 4\kappa (r s \lambda)^{1/2} g_A g_V  {\rm Re} \left(H_{1,V} H_{3,A}^{*}+H_{1,A}H_{3,V}^{*}\right) \\
J_5 &=& \frac{1}{2} \kappa (r s \lambda)^{1/2}  \left(g_A^2+g_V^2\right) {\rm Re} \left(H_{1,V} H_{3,V}^{*}+H_{1,A} H_{3,A}^{*}\right) \\
J_6 &=& 4 (r s)^{1/2} g_A g_V \left[4 \kappa   {\rm Re} \left (H_{1,V}H_{1,A}^{*} \right ) + \lambda   {\rm Re} \left (H_{1,V}H_{2,A}^{*} + H_{1,A}H_{2,V}^{*} \right) \right] \\
J_7 &=& \frac{1}{2} (r s)^{1/2} \left(g_A^2+g_V^2\right) \left[2 \kappa  \left(|H_{1,V}|^{2}+|H_{1,A}|^{2}\right) + \lambda  {\rm Re} \left(H_{1,V} H_{2,V}^{*} + H_{1,A} H_{2,A}^{*} \right) \right] \\
J_8 &=& 2 r s \lambda^{1/2} \left(g_A^2+g_V^2 \right) {\rm Re} \left( H_{1,V} H_{3,V}^{*} + H_{1,A} H_{3,A}^{*} \right) \\
J_9 &=& 2 r s \left(g_A^2+g_V^2\right) \left(|H_{1,V}|^{2}+|H_{1,A}|^{2}\right)
\end{eqnarray}
where to $\mathcal{O}(1/\Lambda^2)$ the $H_i$ are given by
\begin{eqnarray}
H_{1,V} &=& - \frac{2m_H (\sqrt{2} G_F)^{1/2} r}{r-s} g_V \left(1 + \widehat \alpha_1^{\rm eff}-\frac{\kappa}{r} \widehat \alpha_{ZZ} - \frac{\kappa}{2r} \frac{Q_\ell g_{em}  (r-s)}{s g_V} \widehat \alpha_{AZ} \right) \\
H_{1,A} &=& \frac{2m_H (\sqrt{2} G_F)^{1/2} r}{r-s} g_A \left(1 + \widehat \alpha_2^{\rm eff}-\frac{\kappa}{r} \widehat \alpha_{ZZ}\right)\\
H_{2,V} &=& -\frac{2m_H (\sqrt{2} G_F)^{1/2}}{r-s} g_V \left [ 2  
\widehat\alpha_{ZZ}   + \frac{Q_\ell g_{em} (r-s)}{s g_V} \widehat\alpha_{AZ} \right ]  \\
H_{2,A} &=&  \frac{4m_H (\sqrt{2} G_F)^{1/2}}{r-s} g_A  \widehat\alpha_{ZZ} \\
H_{3,V} &=& -\frac{2m_H (\sqrt{2} G_F)^{1/2}}{r-s} g_V  \left [ 2 \widehat\alpha_{Z\widetilde{Z}} +\frac{Q_\ell  g_{em}  (r-s)}{s g_V}  \widehat\alpha_{A\widetilde{Z}}\right] \\
H_{3,A} &=& \frac{4m_H (\sqrt{2} G_F)^{1/2}}{r-s}  g_A \widehat \alpha_{Z\widetilde{Z}}
\end{eqnarray}
with $Q_\ell=-1$ and $\kappa \equiv 1 - r - s$.

Note that only six of the $J_i$ functions are independent, so that e.g. $J_5, J_7, J_9$ can be eliminated in terms of the remaining $J_i$ by the relations

\begin{eqnarray}
J_5 &=& \frac{\kappa}{4 \sqrt{rs}} J_8 \nonumber \\
J_7 &=& \frac{\sqrt{rs}}{2 \kappa} \left( \frac{\kappa^2}{2 rs} J_1 + J_2 \right) \\
J_9 &=& J_1 \nonumber
\end{eqnarray}

\section{Statistical uncertainties of the Asymmetry observables}

In this appendix we derive the statistical uncertainty of an asymmetry observable $A$ as a function of its expectation value $\bar{A}$ and the size of the statistics $N$.  Assuming a fixed number of events $N$, one could divide the events into two sets, with $N_+$ events satisfying some criteria ({\it e.g.} for $\mathcal{A}^{(3)}_\phi$ it is $\cos{\phi}>0$) while $N_-$ events fail to satisfy the same criteria, with $N=N_++N_-$\,.  An asymmetry observable $A$ can be defined as
\begin{equation}
A \equiv \frac{N_+ -N_-}{N_+ +N_-} = \frac{2N_+}{N} -1 \, .  \label{eq:A1}
\end{equation}
Assuming each event has a probability $p$ to be counted into $N_+$, then $N_+$ has a binomial distribution with standard deviation $\sigma_{N_+} = \sqrt{N\, p(1-p)}$, which 
transforms to a standard deviation of $A$ as
\begin{equation}
\sigma_A = 2\sqrt{\frac{p(1-p)}{N}} \,.  \label{eq:sigA}
\end{equation}
The value $p$ is directly related to $\bar{A}$, the expectation value of $A$.  From Eq.~(\ref{eq:A1}) one has $\bar{A} = 2p-1$, which gives
\begin{equation}
p = \frac{1+\bar{A}}{2} \, .    \label{eq:p}
\end{equation}
Combining Eq.~(\ref{eq:sigA}) and Eq.~(\ref{eq:p}), we obtain
\begin{equation}
\sigma_A = \sqrt{\frac{1-\bar{A}^2}{N}} \,.   \label{eq:sigA2}
\end{equation}
In the ideal case with only statistical uncertainty and no background, no systematic error, and with perfect resolution, for the asymmetry observables considered in this paper, $\bar{A}$ is given by the theoretical predictions in Eq.~(\ref{eq:Atheta1n}--\ref{eq:Ac12n}).  The SM expectations and the corresponding uncertainties are listed in Table~\ref{tab:Ath}.  If detector effects are included, Eq.~(\ref{eq:sigA2}) still applies, but $\bar{A}$ has to be modified accordingly to take count of that.  In practice, as long as $\bar{A}$ is not too large, $\sigma_A  \approx 1/\sqrt{N}$ would be a good approximation.  It should be noted that, as discussed in Section~\ref{sec:actual}, the detector effects can affect the sensitivity of the asymmetry observables to new physics, which is not shown in Eq.~(\ref{eq:sigA2}).


\providecommand{\href}[2]{#2}\begingroup\raggedright\endgroup


\end{document}